\documentclass{emulateapj}
\usepackage{float}

\begin{document}

\title[MZ Relation at $\lowercase{z}\sim1.6$]{The FMOS-Cosmos Survey of Star-Forming Galaxies at $\lowercase{z}\sim1.6$ II. The Mass-Metallicity Relation and the Dependence on Star Formation Rate and Dust Extinction}


\author{H.~J.~Zahid\altaffilmark{1},
D.~Kashino\altaffilmark{2}, 
J.~D.~Silverman\altaffilmark{3},
L.J.~Kewley\altaffilmark{4},  
E.~Daddi\altaffilmark{5}, 
A.~Renzini\altaffilmark{6}, 
G.~Rodighiero\altaffilmark{7},
T.~Nagao\altaffilmark{8},
N.~Arimoto\altaffilmark{9,10}, 
D.~B. Sanders\altaffilmark{1}, 
J.~Kartaltepe\altaffilmark{11}, 
S.~J.~Lilly\altaffilmark{12}, 
C.~Maier\altaffilmark{13}
M.~J.~Geller\altaffilmark{14}
P.~Capak\altaffilmark{15,16}, 
C.~M.~Carollo\altaffilmark{12}, 
J.~Chu\altaffilmark{1}, 
G.~Hasinger\altaffilmark{1}, 
O.~Ilbert\altaffilmark{17}, 
M.~Kajisawa\altaffilmark{18}, 
A.~M.~Koekemoer\altaffilmark{19} 
K.~Kova\u{c}\altaffilmark{8}, 
O.~ Le F\`{e}vre\altaffilmark{17}, 
D.~Masters\altaffilmark{20}, 
H.~J.~McCracken\altaffilmark{21}, 
M.~Onodera\altaffilmark{12}, 
N.~Scoville\altaffilmark{22}, 
V.~Strazzullo\altaffilmark{10},
N.~Sugiyama\altaffilmark{2,4}, 
Y.~Taniguchi\altaffilmark{18}
And
The COSMOS Team
}

\email{zahid@cfa.harvard.edu}
\altaffiltext{1}{
Institute for Astronomy, University of Hawaii at Manoa, Honolulu, HI 96822, USA
}
\altaffiltext{2}{
Division of Particle and Astrophysical Science, Graduate School of Science, Nagoya University, Nagoya, 464-8602, Japan
}
\altaffiltext{3}{
Kavli Institute for the Physics and Mathematics of the Universe (WPI), Todai Institutes for Advanced Study, the University of Tokyo, Kashiwanoha, Kashiwa, 277-8583, Japan
}
\altaffiltext{4}{
Research School of Astronomy and Astrophysics, The Australian National University, Cotter Road, Weston Creek, ACT 2611
}
\altaffiltext{5}{
CEA-Saclay, Service d'Astrophysique, F-91191 Gif-sur-Yvette, France
}
\altaffiltext{6}{
INAF Osservatorio Astronomico di Padova, vicolo dell'Osservatorio 5, I-35122 Padova, Italy
}
\altaffiltext{7}{
Dipartimento di Astronomia, Universit\`{a} di Padova, vicolo dell Osservatorio 3, I-35122 Padova, Italy
}
\altaffiltext{8}{
The Hakubi Center for Advanced Research, Kyoto University, Kyoto 606-8302, Japan
}
\altaffiltext{9}{
National Astronomical Observatory of Japan, Subaru Telescope, 650 North A’ohoku Place, Hilo, HI 96720, USA
}
\altaffiltext{10}{
The Graduate University for Advanced Studies, Department of Astronomical Sciences, Osawa 2-21-1, Mitaka, Tokyo,  181-8588, Japan
}
\altaffiltext{11}{
National Optical Astronomy Observatory, 950 N. Cherry Ave., Tucson, AZ, 85719
}
\altaffiltext{12}{
Institute for Astronomy, ETH Z\"{u}rich, Wolfgang-Pauli-strasse 27, 8093 Z\"{u}rich, Switzerland
}
\altaffiltext{13}{
Vienna University, Department of Astrophysics, Tuerkenschanzstrasse 17, 1180 Vienna, Austria
}
\altaffiltext{14}{
Smithsonian Astrophysical Observatory, 60 Garden Street, Cambridge, MA 02138, USA
}
\altaffiltext{15}{
California Institute of Technology, 1200 E. California Blvd., Pasadena, CA 91125, USA
}
\altaffiltext{16}{
Spitzer Science Center, MS 314-6, California Institute of Technology, Pasadena, CA 91125, USA
}
\altaffiltext{17}{
Aix Marseille Universit\'e, CNRS, LAM (Laboratoire d'Astrophysique de Marseille) UMR 7326, 13388, Marseille, France
}
\altaffiltext{18}{
Research Center for Space and Cosmic Evolution, Ehime University, Bunkyo-cho 2-5, Matsuyama, Ehime 790-8577, Japan
}
\altaffiltext{19}{
HST and JWST Instruments/Science Division, Space Telescope Science Institute, 3700 San Martin Drive, Baltimore MD 21218, U.S.A.
}
\altaffiltext{20}{
The Observatories of the Carnegie Institution for Science, 813 Santa Barbara Street, Pasadena, CA 91101, USA
}
\altaffiltext{21}{
Institut d'Astrophysique de Paris, UMR7095 CNRS, Universit\'e Pierre et Marie Curie, 98 bis Boulevard Arago, 75014 Paris, France
}
\altaffiltext{22}{
California Institute of Technology, MC 249-17, 1200 East California Boulevard, Pasadena, CA 91125, USA
}

\begin{abstract}

We investigate the relationships between stellar mass, gas-phase oxygen abundance (metallicity), star formation rate, and dust content of star-forming galaxies at z$\sim$1.6 using Subaru/FMOS spectroscopy in the COSMOS field. The mass-metallicity relation at $z\sim1.6$ is steeper than the relation observed in the local Universe. {The steeper MZ relation at $z\sim1.6$  is mainly due to evolution in the stellar mass where the MZ relation begins to turnover and flatten. This turnover mass is 1.2 dex larger at $z\sim1.6$.} The most massive galaxies at $z\sim1.6$ ($\sim 10^{11}M_\odot$) are enriched to the level observed in massive galaxies in the local Universe. The mass-metallicity relation we measure at $z\sim1.6$ supports the suggestion of an empirical upper metallicity limit that does not significantly evolve with redshift. We find an \emph{anti-}correlation between metallicity and star formation rate for galaxies at a fixed stellar mass at $z\sim1.6$ which is similar to trends observed in the local Universe. We do not find a relation between stellar mass, metallicity and star formation rate that is independent of redshift; our data suggest that there is redshift evolution in this relation. We examine the relation between stellar mass, metallicity and dust extinction. We find that at a fixed stellar mass dustier galaxies tend to be more metal rich. From examination of the stellar masses, metallicities, SFRs and dust extinctions we conclude that stellar mass is most closely related to dust extinction.

\end{abstract}

\section{Introduction}

Near-infrared multi-object spectrographs placed on 8 - 10 m class telescopes have recently opened up the redshift desert ($1<z<2$) for spectroscopic exploration. This redshift range is particularly important since it marks the epoch where galaxies transition from the peak of cosmic star-formation to the more quiescent build-up of stellar mass that we see in galaxies today \citep[][and references therein]{Hopkins2006}. Understanding of the physical processes responsible for this transition is crucial for building a coherent picture of galaxy evolution. In this series of papers, we report on the first results of our recent survey of star-forming galaxies at $1.4<z<1.7$. We present the sample and survey design in Silverman et al. (In Preparation, hereafter Paper III) and the spectroscopically measured, extinction corrected star formation rates (SFRs) in \citet[hereafter Paper I]{Kashino2013}. Here we present on the relation between stellar mass, gas-phase oxygen abundance, SFR and dust extinction for our sample.

The gas-phase oxygen abundance (metallicity) is a crucial diagnostic of galaxy evolution. Oxygen is the most abundant heavy element produced in massive stars and comprises half the mass of heavy elements in the Universe. Therefore, the abundance of oxygen is an excellent proxy of chemical evolution. Oxygen is dispersed into the interstellar medium (ISM) of galaxies by massive stars through stellar winds and supernovae. The mass of oxygen in the ISM accumulates as galaxies build-up stellar mass. However, metallicity is a measure of the amount of oxygen \emph{relative} to hydrogen. Therefore, it is not simply an accumulated record of star-formation but also a sensitive tracer of gas flows. The inflow of pristine gas can dilute the abundance of oxygen and decrease the metallicity but inflows also fuel star-formation leading to the synthesis of heavy elements. At the same time, feedback from massive stars is one of the primary mechanisms by which gas is expelled from galaxies \citep[e.g.][]{Mathews1971, Larson1974}. It is clear that outflows can transport metals out of galaxies \citep{Renzini1997, Martin2002, Kirby2011, Bordoloi2011, Newman2012, Zahid2012b}. However, the impact that outflows have on metallicity remains uncertain since composition of outflowing material is not well constrained observationally. 

Using observations of 8 local star-forming galaxies, \citet{Lequeux1979} first showed that metallicity increases with stellar mass. In subsequent years samples have grown considerably. \citet{Tremonti2004} establish a tight ($\sim0.1$ dex scatter) mass-metallicity (MZ) relation in the local universe by examining $\sim50,000$ galaxies from the SDSS with stellar masses ranging from $10^{8.5} \gtrsim M_\ast/M_\odot \gtrsim 10^{11}$. The relation has since been extended down to $\sim10^6 M_\odot$ \citep{Lee2006, Zahid2012a, Berg2012}. Surveys of distant galaxies have made it possible to study the MZ relation at intermediate \citep{Savaglio2005, Maier2005, Zahid2011a, Perez-Montero2013, Zahid2013b} and high redshifts \citep{Erb2006b, Maiolino2008, Mannucci2009, Laskar2011, Yabe2012, Yuan2013, Kulas2013}. At a fixed stellar mass, galaxies are less enriched at higher redshifts. While the origin of this relation is still debated, measurements of the chemical evolution of galaxies provide important constraints for the processes of star formation and gas flows in models of galaxy evolution \citep[e.g.,][]{Brooks2007, Finlator2008, Dave2011b, Zahid2012b, Torrey2013, Lilly2013, Zahid2014a}.

While almost all studies conclude that the gas in galaxies becomes more metal-rich as the Universe evolves, some studies also report a flattening of the the MZ relation for massive galaxies at late times \citep{Savaglio2005, Maier2005, Maiolino2008, Zahid2011a, Zahid2013b}. Many of these works \citep{Savaglio2005, Maier2006, Maiolino2008, Zahid2011a} attribute this flattening to galaxy downsizing \citep{Cowie1996}, i.e. the process by which star formation becomes more dominant in lower mass systems at late times. However, we show that flattening of the slope of the MZ relation is more consistent with the process of metallicity saturation rather than strictly a consequence of downsizing \citep{Zahid2013b}. In this study, we revisit this issue and extend our analysis to $z\sim1.6$.

In the local Universe, the metallicity at a fixed stellar mass appears to be correlated with SFR \citep{Ellison2008, Lara-Lopez2010, Mannucci2010, Yates2012, Andrews2013}. \citet{Mannucci2010} suggest that the relation between stellar mass, metallicity and SFR that minimizes the scatter for local galaxies does not evolve out to $z\sim2.5$. However, several studies have shown that this relation is dependent on methodology \citep{Yates2012, Andrews2013}. We examine the relation between stellar mass, metallicity and SFR at $z\sim1.6$ applying a consistent methodology throughout.

The paper is organized as follows: In Section 2 and 3 we describe our data and methodology, respectively. In Section 4 we examine emission line diagnostics using a subset of our sample where we have both $J$ and $H$-band observations. In Section 5 we present the main results of our study and we discuss potential systematic issues in our measurements in Section 6. In Section 7 we discuss our results and we present a summary in Section 8. When necessary we adopt a standard cosmology with $(H_{0}, \Omega_{m}, \Omega_{\Lambda}) = (70$ km s$^{-1}$ Mpc$^{-1}$, 0.3, 0.7) and a \citet{Chabrier2003} IMF.

\section{Data}

\subsection{FMOS-COSMOS Observations}

\begin{figure*}
\begin{center}
\includegraphics[width=2\columnwidth]{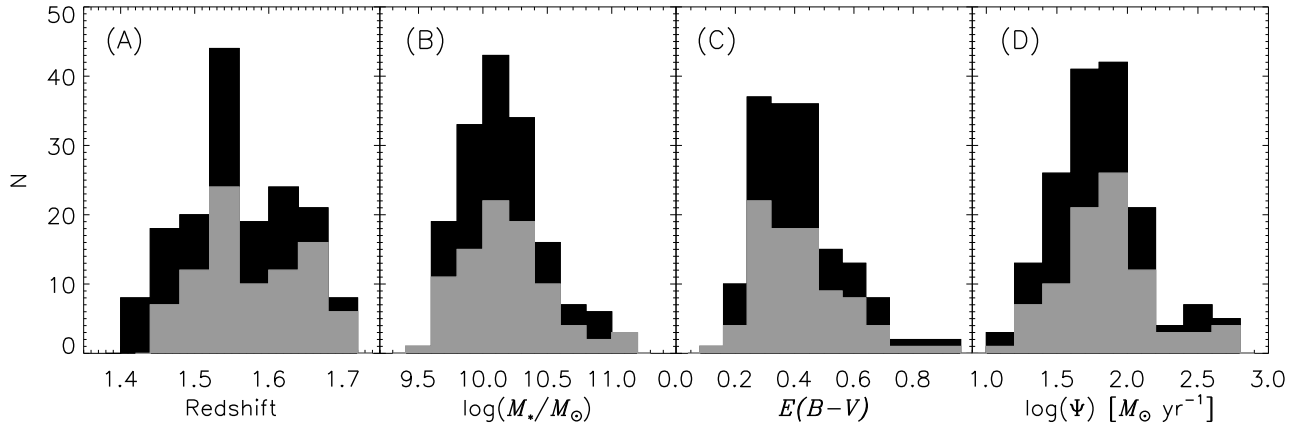}
\end{center}
\caption{The (A) redshift, (B) stellar mass, (C) nebular $E(B-V)$ and (D) SFR distribution of our sample of galaxies at $z\sim1.6$. The black histogram shows the distribution for our sample of 162 H$\alpha$ detected galaxies. The gray histogram shows the distribution for the subsample of 85 H$\alpha$ detected galaxies that are observed both in $J$ and $H$-band.}
\label{fig:hist}
\end{figure*}

Details of the survey design and observations are presented in Paper III. Here we summarize the most relevant aspects. We emphasize that when necessary measured quantities are converted from the \citet{Salpeter1955} IMF used in Paper I and III to the \citet{Chabrier2003} IMF used in this work. This is done for consistency with our previous metallicity studies.

Our observations are carried out using the near-infrared Fiber Multi-Object Spectrograph \citep[FMOS;][]{Kimura2010} on the Subaru Telescope. FMOS has 400 1.2'' fibers distributed over a 30' diameter circular field-of-view. We operate the spectrograph in cross-beam switching mode. Two fibers are allocated to each object. The spectrograph dithers between two positions such that one of the two fibers is always on source while the other fiber is used for sky subtraction. This procedure allows us to observe $\sim200$ galaxies with simultaneous sky observations for optimal sky subtraction. An OH-airglow suppression filter blocks the strongest atmospheric emission lines \citep{Iwamuro2001}. Our observations are taken using the high resolution mode which has a spectral resolution of $R\sim2200$. At this resolution, the [NII]$\lambda6584$ and H$\alpha$ lines are well resolved in star-forming galaxies \citep[see Figure 1 in][]{Kashino2013} and contamination from narrow sky lines is minimized. For galaxies at $1.4<z<1.7$, we can observe H$\alpha$ and H$\beta$ in the $H$-long (1.60 - 1.80 $\mu$m) and $J$-long (1.11 - 1.35 $\mu$m) bands, respectively.

We primarily target star-forming galaxies in the redshift range of $1.4<z<1.7$ in the central square degree of the COSMOS field \citep{Scoville2007, Koekemoer2007}. We preselect galaxies using robust photometric redshifts from the catalog of \citet{Ilbert2009}. These redshifts are based on 30 bands of photometry ranging from the UV to the mid-infrared. In order to efficiently target star-forming galaxies, we require $K_s$-band magnitudes $<23$. For the majority of the sample we use an sBzK selection \citep{Daddi2004} using the catalog of \citet{McCracken2010}. A significant number of objects however, were selected based on stellar mass and photo-z (see Paper III). To minimize AGN contamination, we exclude galaxies which have x-ray detections. We revisit this issue in Section 4 and 6. 

The $H$-band observations of 796 galaxies were carried out over 6 nights in March 2012 and 2 nights in January 2013. Each galaxy was observed for $\sim5$ hours allowing us to reach a 3$\sigma$ flux limit of $4 \times 10^{-17}$ ergs s$^{-1}$ cm$^{-2}$ corresponding to a unobscured SFR limit of $\sim5 M_\odot$ yr$^{-1}$. A subsample of galaxies with H$\alpha$ detections satisfying signal-to-noise (S/N) $>3$  were observed in the $J$-band in order to obtain measurements of H$\beta$ and [OIII]$\lambda5007$. These observations were carried out in March 2012, December 2012 and February 2013 with an on source integration time of $\sim5$ hours. 

All data were reduced with the FMOS Image-Based Reduction Package \citep[Fibre-pac;][]{Iwamuro2012}. In cross-beam switching mode, the sky is observed simultaneously along with the target. Initial sky subtraction is performed by differencing the fiber pair. Next the detector cross talk and bias difference between the quadrants are corrected and the flat fielding is performed. After the bad pixels are rejected, the image is corrected for distortion and the residual sky is subtracted. The images from multiple exposures are combined to produce a single spectrum per object. The wavelength is calibrated based on the Th-Ar spectral images taken just before the science exposure with a typical accuracy of $>1$ pixel ($\sim1$\AA ~ in high resolution mode). The flux calibration is carried out using flux standard stars that are observed simultaneously with other scientific targets.

From our observations we selected 168 galaxies with significant H$\alpha$ detections (S/N $>3$). A subsample of 89 galaxies also have corresponding $J$-band observations. For the metallicity analysis we remove galaxies whose H$\alpha$ emission is near the edge of the detector and therefore [NII]$\lambda6584$ is not observed. Our final sample consists of 162 star-forming galaxies observed in the $H$-band band with significant (S/N $> 3$) detections of H$\alpha$ emission and a subsample of 87 galaxies with corresponding $J$-band observations. The measured physical properties of both these samples are shown in Figure \ref{fig:hist}. Figure \ref{fig:hist} shows that the subsample of galaxies for which we have $J$-band observations is representative of the larger $H$-band sample. 

\subsection{The Local Sample}

{We derive the local MZ relation using data from the SDSS Legacy Survey (SDSS I-II). The latest release of the Legacy data is found in the DR8 \citep{Aihara2011}. The spectroscopic data consists of $\sim900,000$ galaxies spanning a redshift range of $0<z<0.7$. The survey has a limiting magnitude of $r = 17.8$ and covers 8000 deg$^2$ on they sky \citep{Strauss2002}. The nominal spectral range of the observations is $3800 - 9200\mathrm{\AA}$ with a spectral resolution of $R = 1800-2000$. We use the most recent $ugriz-$band c-model magnitudes released as part of the DR8 \citep{Padmanabhan2008}. We use the latest emission line flux measurements released by the Portsmouth Group\footnote{https://www.sdss3.org/dr10/spectro/galaxy\_portsmouth.php} \citep{Thomas2013}. The line fluxes are corrected for dust extinction using the correction equation from \citet{Calzetti2000} and only the corrected fluxes are given in the catalog. In order to derive dust extinction from the Balmer decrement, we use the MPA/JHU group catalog\footnote{http://www.mpa-garching.mpg.de/SDSS/DR7/} of line flux measurements which are not corrected for dust extinction.}

We select galaxies in a limited redshift range with $z<0.12$ to minimize evolutionary effects and we require an aperture covering fraction $\gtrsim20\%$ to avoid biasing our metallicity estimate \citep{Kewley2005}. {\citet{Ellison2008} show that metallicities are correlated to galaxy size and therefore aperture effects may bias the MZ relation. However, the MZ relation we derive does not strongly depend on the minimum covering fraction we apply in selecting the sample. This could result form the fact that massive galaxies, which are most affected by aperture effects, tend to have shallow metallicity gradients. Integral field surveys of local galaxies currently underway will definitely address systematic biases related to covering fractions.}

{Our primary selection criteria for galaxies is the S/N ratios of strong emission lines. The results presented in this paper are based on analysis of the H$\beta$, [OIII]$\lambda5007$, H$\alpha$ and [NII]$\lambda6584$ emission lines. \citet{Foster2012} show that the MZ relation is insensitive to the S/N threshold adopted for the H$\beta$, H$\alpha$ and [NII]$\lambda6584$ emission lines. However, they also show that S/N cuts on the [OIII]$\lambda5007$ line can lead to a significant bias in the MZ relation. We require a S/N $>3$ in the H$\beta$, H$\alpha$ and [NII]$\lambda6584$ emission lines but make no S/N cut on the [OIII]$\lambda5007$ line.}


Metallicities are determined from line flux ratios under the assumption that massive stars produce the EUV radiation field that ionized the nebular gas. Therefore, metallicities determined in galaxies where active galactic nuclei (AGN) contribute significantly to the ionizing radiation are not reliable. AGN are removed from the sample using the [OIII]/H$\beta$ vs. [NII]/H$\alpha$ line flux ratio diagram \citep[i.e. the BPT method,][]{Baldwin1981, Kauffmann2003, Kewley2006}. Star-forming galaxies are well separated from AGN in the [OIII]/H$\beta$ vs. [NII]/H$\alpha$ line flux ratio diagram and her we apply the separation given in \citet{Kewley2006} to remove AGN. Galaxies with 
\begin{equation}
\mathrm{log([OIII]/H}\beta) < 0.61/[\mathrm{log([NII]/H}\alpha) - 0.05] + 1.3
\end{equation}
are defined as star-forming. Here, [OIII], H$\beta$, [NII] and H$\alpha$ are the emission line strengths of [OIII]$\lambda5007$, H$\beta$, [NII]$\lambda6584$ and H$\alpha$, respectively. 

Our final sample consists of $\sim88,000$ star-forming galaxies in the redshift range of $0.02<z<0.12$. 

\section{Methods}

\subsection{Metallicity Determination}

Collisionally excited emission lines are the primary coolant in HII regions. Their line strengths scale with temperatures and metallicity. Therefore, flux ratios of collisionally excited emission lines to recombination emission lines can be used to estimate metallicity. Various ratios have been calibrated either empirically or theoretically to yield estimates of the gas-phase oxygen abundance. Different calibrations applied to the same galaxies can give metallicities that range up to $\sim0.6$ dex \citep{Kewley2008}. Thus, there is great uncertainty in the absolute metallicity scale with empirically calibrated diagnostics typically yielding metallicities that are $\sim0.3$ dex smaller than theoretically calibrated diagnostics. It is beyond the scope of this paper to go into details of various diagnostics and possible resolutions. We refer the reader to detailed discussions of this issue presented in \citet{Kewley2008}. However, we note that despite the uncertainty in absolute metallicities, \citet{Kewley2008} find that relative metallicities determined using various diagnostics are robust. This is fortunate since we are able to apply the same method for determining metallicity in both our local sample from SDSS and our sample of galaxies at $z\sim1.6$. 

We determine metallicities using the commonly used $N2$ diagnostic calibrated by \citet[PP04 hereafter]{Pettini2004}. Here, $N2$ is defined as the log of the emission line ratio between [NII]$\lambda6584$ and H$\alpha$. The advantage of this ratio is that the lines are closely spaced in wavelength and therefore can be easily observed simultaneously in the $H$-band for galaxies at $1.4<z<1.7$. Additionally, because we are taking a flux ratio and the lines are only separated by $\sim20$\AA, no extinction correction is required. PP04 calibrate the line ratio using a semi-empirical approach. At lower metallicities they empirically determine metallicities from temperature sensitive auroral lines. This method is known as the ``direct" method and it provides a well established metallicity scale below solar metallicities. However, the [OIII]$\lambda4363$ auroral line used in determining metallicities with the ``direct" method is extremely weak. Because the line strength decreases exponentially with increasing metallicity it is not observed in metal-rich HII regions. PP04 use photoionization modeling of individual HII regions to calibrated metallicities in metal-rich regions where the direct method can not be applied. We apply the linear calibration given by 
\begin{equation}
\mathrm{12 + log(O/H)} = 8.90 + 0.57 \times N2.
\end{equation}
The calibration is valid for $-2.5 < N2 < -0.3$. The formal statistical errors of the slope and intercept are 0.03 and 0.04, respectively, and the intrinsic dispersion is 0.18 dex. {We note that PP04 also provide a quadratic calibration of the N2 ratio. While the quadratic calibration gives a quantitatively different MZ relation, the major conclusions of this work rely on the relative accuracy of the diagnostics and are independent of the particular calibration. Throughout this work, we provide the measured N2 values along with the inferred metallicities.}

PP04 also calibrate the $O3N2$ line ratio which is defined as log\{([OIII]/H$\beta$)/([NII]/H$\alpha$)\}. Here, [OIII] refers to the line flux of [OIII]$\lambda5007$ and [NII] refers to the line flux of [NII]$\lambda6584$. The metallicity calibration for this ratio is
\begin{equation}
\mathrm{12 + log(O/H)} = 8.73 - 0.32 \times O3N2.
\end{equation}
Because the $N2$ and $O3N2$ line ratios are calibrated using the same data, they should be consistent. However, several studies of high redshift galaxies have reported systematic differences in the metallicities determined using this ratio \citep[e.g.][]{Erb2006b, Yabe2012}. Since we measure the [OIII] and H$\beta$ lines in a subset of our galaxies, we are able to asses the consistency of these two ratios at $z\sim1.6$ (see Section 4.2).

Several authors have calibrated the $N2$ ratio independently \citep{Denicolo2002, Nagao2006, Marino2013}. While the calibrations vary systematically, they are all consistent with a monotonically increasing, quasi-linear relation between metallicity and $N2$. We emphasize that while the absolute metallicity is uncertain and varies systematically depending on the calibration applied, the relative metallicities determined from $N2$ are robust over the range of metallicities explored in this study. The results and conclusions of this study rely on the relative metallicities being accurate. In Table \ref{tab:data} we provide the measured $N2$ ratio and the metallicities derived using the PP04 calibration.

\subsection{Mass Determination}

For the SDSS sample we determine the stellar masses from the $ugriz$-band photometry \citep{Stoughton2002}. The stellar masses for the FMOS sample are determined from 30 band UV to IR photometry \citep{Capak2007}.

We use the Le Phare\footnote{$\url{http://www.cfht.hawaii.edu/{}_{\textrm{\symbol{126}}}arnouts/LEPHARE/cfht\_}$ $\url{lephare/lephare.html}$} code developed by Arnouts, S. \& Ilbert, O. to estimate stellar masses. We estimate the stellar masses of galaxies by comparing photometry with stellar population synthesis models in order to determine the mass-to-light (M/L) ratio. The M/L ratio is then used to scale the observed luminosity \citep{Bell2003b, Fontana2004}. Magnitudes are synthesized from the stellar templates of \citet{Bruzual2003} and we use a \citet{Chabrier2003} IMF. The 27 models have two metallicities and seven exponentially decreasing star formation models (SFR $\propto e^{-t/\tau}$) with $\tau = 0.1,0.3,1,2,3,5,10,15$ and $30$ Gyrs. We apply the extinction law of \citet{Calzetti2000} allowing E(B$-$V) to range from 0 to 0.6 and the stellar population ages range from 0 to 13 Gyrs. \citet{Conroy2009a} estimate that systematic errors in stellar mass are $\sim0.3$ dex for local galaxies. We have applied a consistent procedure for measuring the stellar masses for our different samples in order to mitigate systematic uncertainties and derive a relatively robust estimate.

We use the \citet{Kennicutt1998b} relation between the synthesized UV luminosity and SFR to correct for the emission line contributions to the photometry. This treatment accounts for H$\alpha$, H$\beta$ and [OII]$\lambda3727$ and [OIII]$\lambda4959, 5007$ \citep{Ilbert2009}. We have determined stellar masses using broadband photometry and  therefore the emission line correction are small. In \citet{Zahid2011a} we compare this method with the method used by the MPA/JHU group to determine stellar masses of the SDSS galaxies. We find that the dispersion between the two methods is 0.14 dex which is consistent with the observational uncertainties. The stellar mass distribution of our sample is plotted in Figure \ref{fig:hist}B.

\subsection{E(B-V) Determination}

For the local sample we measure dust extinction from the Balmer decrement. For case B recombination with electron temperature T$_e$  = 10$^4$K and electron density $n_e = 10^2$ cm$^{-3}$, the intrinsic H$\alpha$/H$\beta$ ratio is expected to be 2.86 \citep{Hummer1987}. We obtain the intrinsic color excess, $E(B-V)$ using the extinction curve of \citet{Calzetti2000}. \citet{Groves2012} suggest that H$\beta$ equivalent widths and line fluxes provided in the SDSS DR7 are underestimated due to improper correction for H$\beta$ absorption. We apply the correction they recommend. The typical correction is small, reducing the $E(B-V)$ by $\sim0.03$.

We detect H$\beta$ in very few individual galaxies at $z\sim1.6$. Therefore, we are not able to measure extinction from the Balmer decrement in individual galaxies. {In Paper I we determine stellar reddening, $E_{star}(B-V)$, from the observed $B_J - z$ color \citep{Daddi2007}. We average our FMOS-COSMOS spectra in three bins of $E_{star}(B-V)$. We measure the H$\beta$ and H$\alpha$ line fluxes and derive the Balmer decrement and nebular reddening, $E_{neb}(B-V)$, from these average spectra. We determine the average factor $f = 0.76$ which relates stellar reddening to nebular reddening, i.e. $E_{neb}(B-V) = E_{star}(B-V)/f$. We convert the stellar reddening determined from the observed $B_J - z$ color of individual galaxies into a nebular reddening, $E_{neb}(B-V)$, using our correction factor, $f$.} The extinction correction, $A_{\mathrm{H}\alpha}$, ranges between $0.6\sim1.7$. Hereafter, all references to $E(B-V)$ are to $E_{neb}(B-V)$ derived from the stellar extinction using the $f=0.76$ factor. The $E(B-V)$ distribution is plotted in Figure \ref{fig:hist}C.

\subsection{SFR Determination}

For the SDSS sample \citet{Brinchmann2004} measure SFRs from the Balmer lines but apply an aperture correction based on the galaxy colors measured inside and outside the fiber. \citet{Salim2007} improve this correction by comparing SFRs determined from Balmer lines with SFRs determined from UV photometry. We emphasize that aperture corrections to the SFR are important since the 3'' SDSS fibers typically cover less than half of the galaxy light \citep{Kewley2005, Zahid2013a}. The aperture corrected SFRs are made available by the MPA/JHU group in the DR7 and we adopt them in this work. We convert the SFRs to a Chabrier IMF by subtracting 0.05 dex from the DR7 measurements.

We measure SFRs for galaxies at $z\sim1.6$ from extinction corrected H$\alpha$ luminosities using the calibration from \citet{Kennicutt1998b}. The measurements in Paper I are based on the \citet{Salpeter1955} IMF. We scale down these measurements by a factor of 1.7 for consistency with the \citet{Chabrier2003} IMF we use in this work. The cosmology adopted in Paper I is $\Omega_m = 0.25$ and $\Omega_\Lambda = 0.75$. This differs from the cosmology adopted in this work ($\Omega_m = 0.3$, $\Omega_\Lambda = 0.7$). We scale the SFRs used in this work down by a factor of 0.035 dex to convert between the two cosmologies. The FMOS fibers are 1.2" and therefore do not typically fully cover most galaxies at $z\sim1.6$ under typical seeing conditions. We correct for fiber losses by applying an aperture correction to the data. {The aperture correction is derived by convolving \emph{HST/ACS} $I$-band images \citep{Koekemoer2007} with the seeing conditions of our observation and extracting 1.2'' aperture flux and comparing it to total flux. This procedure assumes that the H$\alpha$ and UV emission have the same spatial distribution. The H$\alpha$ flux is scaled by the aperture correction factor which ranges between $1.2\sim5$ with a typical value of $\sim2$.}

The SFR distribution of our sample is shown in Figure \ref{fig:hist}D.

\subsection{Averaging Spectra}

In star-forming galaxies the [NII] emission line is significantly weaker than H$\alpha$. Because the line scales with metallicity, it is more likely to be detected in metal-rich galaxies and is detected in only a small fraction of our sample. In order to derive an unbiased MZ relation, we rely on stacking multiple spectra sorted by stellar mass in order to measure an average [NII]/H$\alpha$ ratio. We find that [NII] can be measured with reasonable S/N in even the least massive (lowest metallicity) galaxies in our sample by stacking $\sim16$ spectra.

{We tried various methods for stacking the data including error weighted averages, medians and stacking of spectra normalized to luminosity or SFR. In each case, the MZ relation we derive was statistically consistent. We adopt the procedure described below because it yields the smallest $\chi^2$ fit of the MZ relation.}

\begin{figure*}
\begin{center}
\includegraphics[width=2\columnwidth]{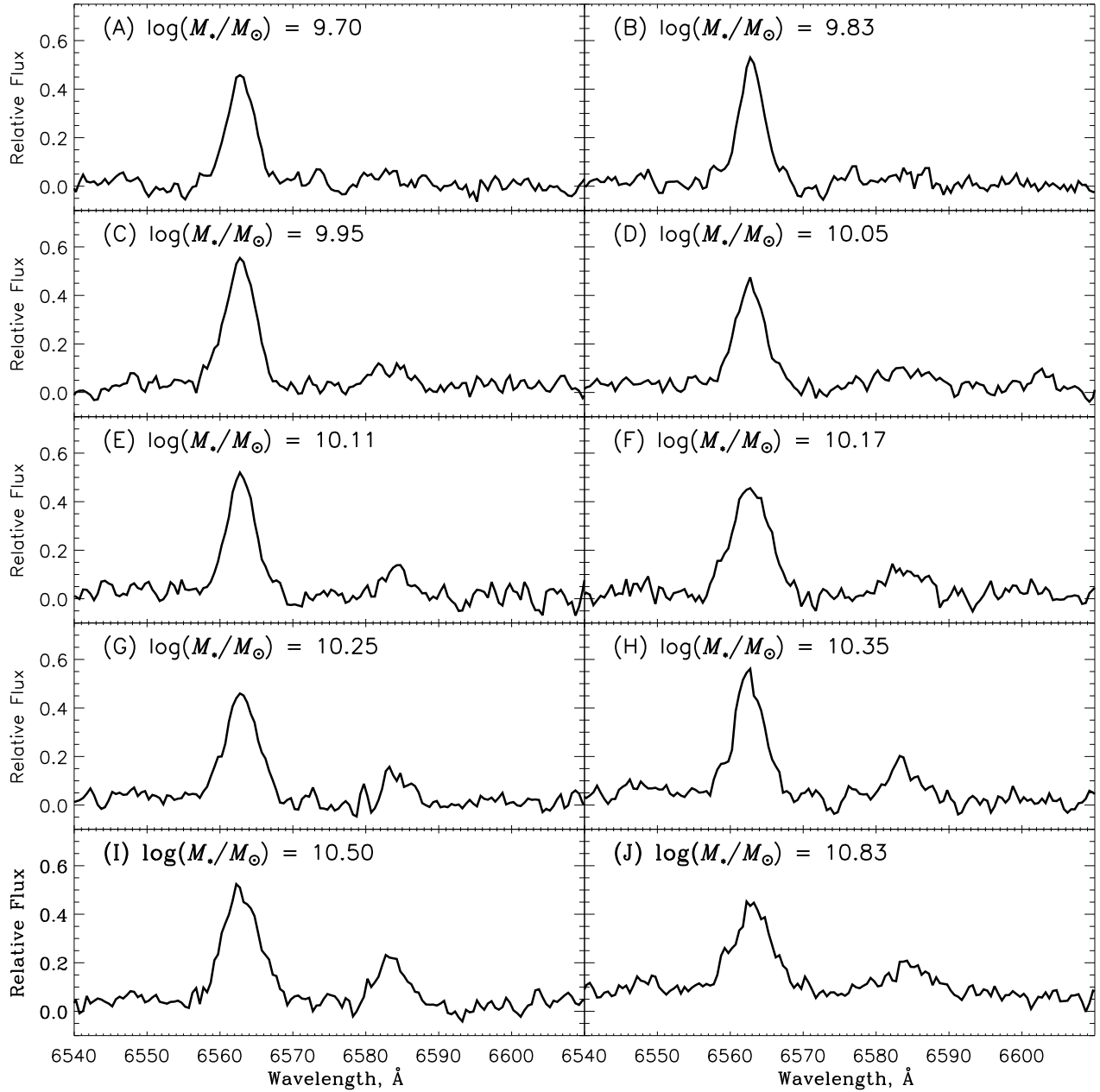}
\end{center}
\caption{The FMOS spectra sorted and averaged in 10 stellar mass bins. The median stellar mass of each bin is listed in each panel.}
\label{fig:spectra}
\end{figure*}

We sort galaxies into bins of stellar mass and average the spectra. We convert the observed frame flux vector of each observation to the rest-frame flux vector using the measured redshift. We then interpolate the flux and error spectrum of each galaxy using a 0.5\AA ~per pixel sampling. This wavelength sampling corresponds to the observed frame single pixel resolution of FMOS (1.25\AA) for galaxies at $z=1.5$. Before averaging over the multiple observations we perform two cuts. First, we remove pixels in regions contaminated by strong residual sky lines. These regions are easily identified in the error spectrum by their relatively large errors. This cut typically removes $\sim10\%$ of the data in each resampled pixel bin. Second, we calculate a resistant mean using \emph{resistant\_mean.pro} routine in IDL which is part of the robust statistics package in the Astronomy Users Library. We clip all data that is 5 median absolute deviations from the median of the distribution of each pixel. This procedure typically cuts out $\sim2\%$ of the data. Our results are not sensitive to the specific level adopted for the two cuts. The error of each resolution element in the average is determined from the observational uncertainties of the individual pixels in each spectra added in quadrature. Figure \ref{fig:spectra} shows the stacked spectra.

A serious concern is that the data cuts we have applied may bias the MZ relation we derive in Section 5.1. In order to assess the impact of our cuts on the MZ relation we apply each of the cuts described above in turn. The MZ relation we derive by averaging the data without applying any cuts is statistically consistent with relation we derive with the cuts applied. The average error in the metallicity we measure from the average spectra when no cuts are applied is $\bar{\sigma} = 0.11$ dex and the $\chi^2$ of the fit to the MZ relation is $2.9$. By removing $\sim10\%$ of the data in regions contaminated by strong sky lines we reduce the average error in the metallicity to $\bar{\sigma} = 0.036$ dex and derive an improved fit with $\chi^2 = 2.4$. Finally, by using the resistant mean and removing $\sim2\%$ of the data the average errors remain the same but the derived fit is improved to $\chi^2 = 1.6$. We conclude that the procedure we apply in removing data does not bias our derived relation but does significantly reduce the errors in metallicity and improve the fit of the MZ relation. 

\subsection{Line Fitting}

We fit emission lines using the \emph{MPFIT} package of routines implemented in IDL \citep{Markwardt2009}. The [NII]$\lambda6584$ and H$\alpha$ line are simultaneously fit with a gaussian profile. We first subtract away the continuum by fitting a line to the pixels in 40\AA~ windows on either side of the emission lines. We then perform a $\chi^2$ minimization to fit the width, amplitude and central wavelength of each emission line. The line widths of the [NII]$\lambda6584$ and H$\alpha$ emission lines are forced to be the same. We adopt the area of the gaussian determined from the fit parameters as our estimate of the line flux. We propagate the observational uncertainties to the fit parameters and flux estimates. 

For a subset of our galaxies we have $J$-band observations of the [OIII]$\lambda5007$ and H$\beta$ emission lines. We follow an identical procedure for estimating line fluxes for these lines as we do for [NII] and H$\alpha$ but with one notable exception. Balmer absorption arising in the atmospheres of intermediate-mass A-type stars can lead to underestimates of the H$\beta$ emission line flux. This is because the emission line sits in an absorption trough which leads to a flux decrement \citep[e.g., see Figure 3 in][]{Zahid2011a}. We correct for this underlying absorption by fitting the continuum using stellar population synthesis models. Here we combine our $J$-band and $H$-band observations. We average our 87 spectra in five stellar mass bins. We mask out emission lines and fit the continuum in the average spectra with a linear combination of \citet{Bruzual2003} models convolved to the FMOS instrument resolution. 

\begin{figure}
\begin{center}
\includegraphics[width=\columnwidth]{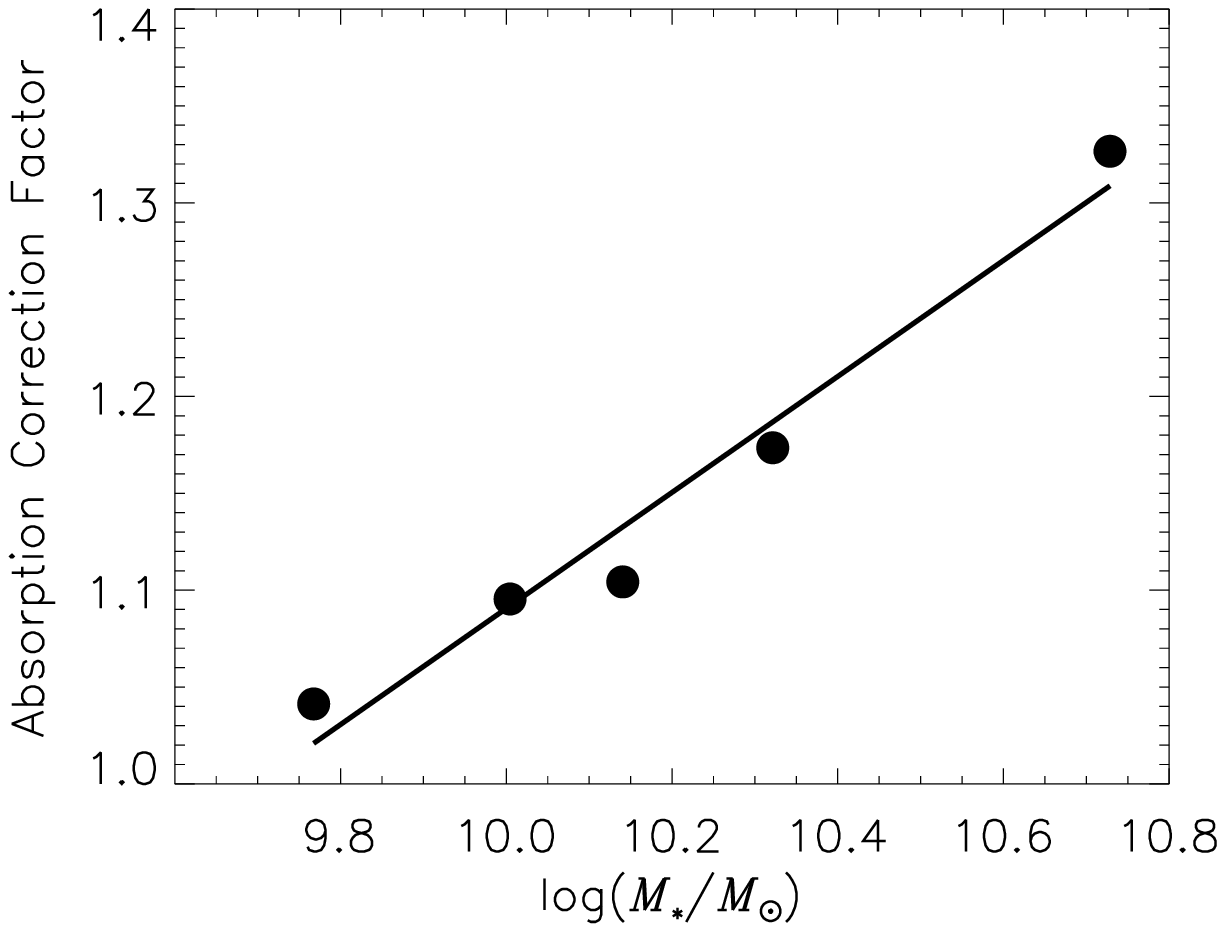}
\end{center}
\caption{The factor by which the H$\beta$ line flux corrected for Balmer absorption is greater than the H$\beta$ line flux without an absorption correction. The factor is determined from average galaxy spectra sorted in five bins of stellar mass. }
\label{fig:abs_corr}
\end{figure}

Figure \ref{fig:abs_corr} shows the Balmer absorption correction factor derived by comparing the fluxes measured with and without an absorption correction. The absorption correction factor is defined as $F(\mathrm{H}\beta)_{corr}/F(\mathrm{H}\beta)$. Here $F(\mathrm{H}\beta)_{corr}$ and $F(\mathrm{H}\beta)$ are the H$\beta$ line fluxes measured from our stacked data with and without an absorption correction, respectively. The linear fit is given by
\begin{equation}
F(\mathrm{H}\beta)_{corr}/F(\mathrm{H}\beta) = 1.09 + 0.30\left[\mathrm{log}(M_\ast/M_\odot) - 10\right]
\label{eq:abs_corr}
\end{equation}
The H$\beta$ emission line Balmer absorption correction ranges from $1\sim1.5$ for galaxies in our sample with a median of 1.27. For the BPT analysis in the following section, we apply the absorption correction given by Equation \ref{eq:abs_corr} to individual galaxies. {The absorption correction for H$\alpha$ is small ($\lesssim2\%$, see Paper I). Thus, we make no correction for H$\alpha$ absorption.}

We derive the formal errors by propagating the observational uncertainties of each pixel through to the fit parameters from which we determine line fluxes. These errors are then propagated to the line ratios and metallicities in the standard way. The formal errors only reflect the quality of the data and do not account for the intrinsic scatter in the line flux ratios due to the intrinsic scatter in the MZ relation \citep{Zahid2012a}. Within each stellar mass bin we estimate the intrinsic scatter in the metallicities by bootstrapping the data. In each bin of stellar mass, we randomly select, with replacement, $N$ spectra and average using the procedure described above. Here $N$ is the number of spectra in each bin of stellar mass (see Table \ref{tab:data}). We perform this procedure 1000 times for each of the mass bins, determining the line fluxes and metallicities from the average spectra each time. Given the small number of spectra in each mass bin ($\sim16$), it is unlikely that we are sampling the full scatter in line fluxes and metallicities. The scatter derived using the bootstrap therefore is a lower limit to the intrinsic scatter.

\section{Emission Line Diagnostics}

\subsection{AGN Contamination}
\begin{figure*}
\begin{center}
\includegraphics[width=1.8\columnwidth]{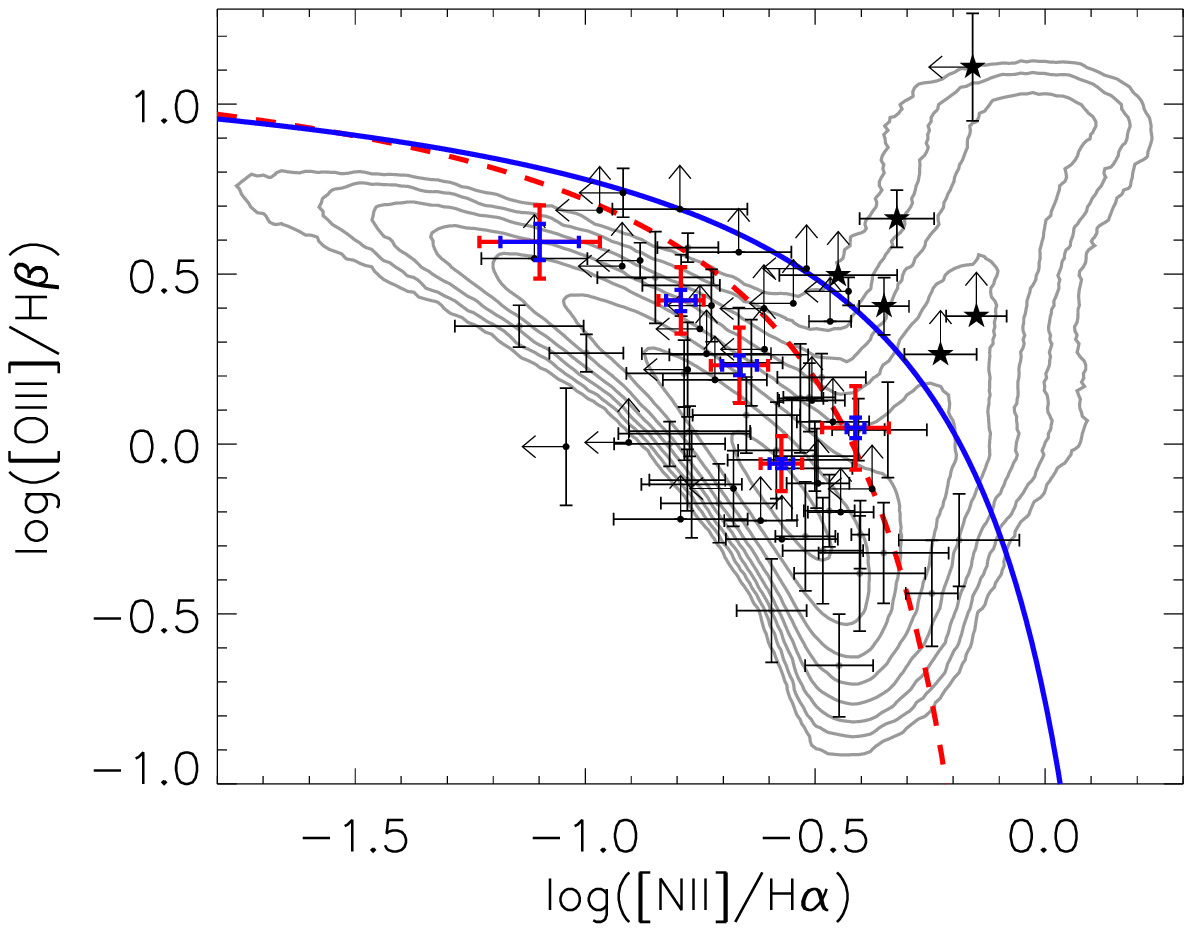}
\end{center}
\caption{The [OIII]/H$\beta$ vs [NII]/H$\alpha$ diagram for the subsample of our galaxies with both $J$ and $H$-band observations. Individual galaxies are shown by black. The objects denoted by stars are identified as AGN by their location on this diagram. We also plot the line fluxes determined from spectra averaged in five mass bins. The blue and red errors bars are the formal and bootstrapped errors, respectively. The gray contours are the distribution of $\sim93,000$ local galaxies in the SDSS. The red curve is the empirical separation between purely star-forming galaxies and composites and AGN in the local Universe \citep{Kewley2006}. The blue curve is the separation at $z\sim1.6$ \citep{Kewley2013b}.}
\label{fig:bpt}
\end{figure*}

The [OIII]/H$\beta$ vs. [NII]/H$\alpha$ diagnostic diagram is commonly used to classify galaxies as star-forming or active galactic nuclei \citep[AGN][]{Baldwin1981, Kauffmann2003, Kewley2006}. The line flux of [OIII]$\lambda5007$ relative to H$\beta$ is plotted against the line flux of [NII]$\lambda6584$ relative to H$\alpha$. Recently, \citet{Kewley2013a} suggest that the physical conditions of the ISM and radiation field evolve with redshift, thus leading to a cosmic evolution of the locus of star-forming and AGN galaxies on the [OIII]/H$\beta$ vs. [NII]/H$\alpha$ diagnostic diagram. By comparing their theoretical models to observations, \citet{Kewley2013b} derive a redshift dependent classification given by
\begin{eqnarray}
\mathrm{log}([OIII]/\mathrm{H}\beta) & = &\frac{0.61}{\mathrm{log}([NII]/\mathrm{H}\alpha) - 0.02 - 0.1833z} \nonumber \\
& & + 1.2 + 0.03z.
\label{eq:bpt}
\end{eqnarray}
Galaxies below and above the dividing line defined by Equation \ref{eq:bpt} are classified as star-forming and AGN, respectively.

In Figure \ref{fig:bpt} we plot the [OIII]/H$\beta$ vs. [NII]/H$\alpha$ line ratios for 72 individual galaxies with H$\alpha$ and [OIII]$\lambda5007$ detections. We also plot the ratios for average spectra of 87 galaxies sorted into 5 mass bins. The emission line properties are determined from averaging 16 or 17 spectra in each stellar mass bin. The blue error bars are the formal observational uncertainties and the red error bars are the bootstrapped errors which estimate the intrinsic scatter of the data. The red dashed and blue solid curves are the local and $z\sim1.6$ classification determined from Equation \ref{eq:bpt}, respectively. We find that our sample of galaxies have excitation that is nearly consistent with the locus of local star-forming galaxies (cf. contours and red dashed line). Comparison with the $z\sim1.6$ classification line reveals that most of our galaxies lie well within the star-forming sequence. The six galaxies denoted by stars are identified as AGN. For our sample we find that $8\% \pm 3\%$ of the sample is AGN. This value is consistent with the estimates of \citet{Stott2013} for galaxies at $z\sim1.5$. 

\begin{figure}
\begin{center}
\includegraphics[width=\columnwidth]{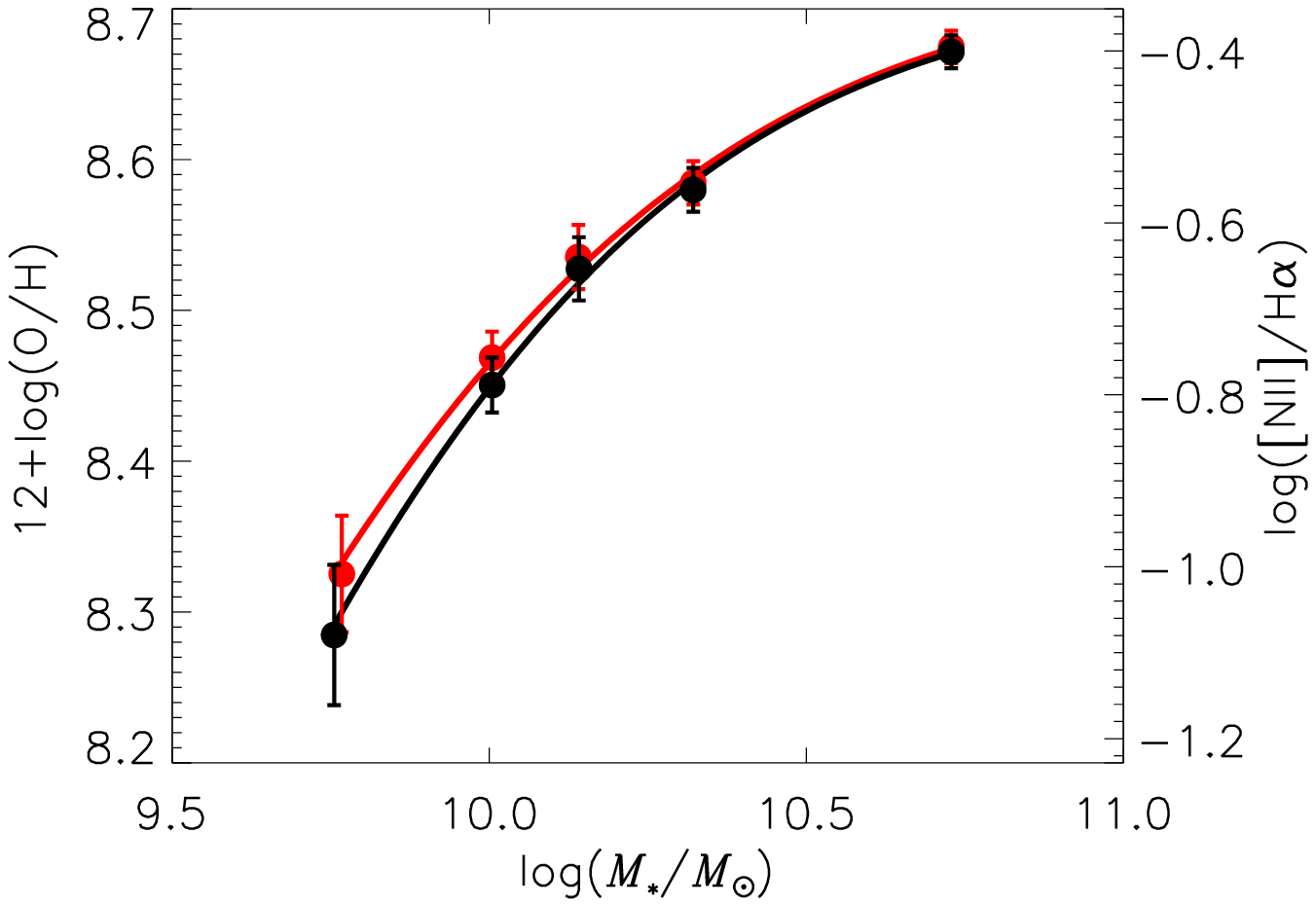}
\end{center}
\caption{A comparison of the MZ relation for our sample of 87 galaxies with $J$ and $H$-band observations. The red curve is determined from the full sample and the black curve is determined by from a subsample of 81 galaxies where the six galaxies identified as AGN in Figure \ref{fig:bpt} are removed. The impact of AGN contamination appears to be negligible.}
\label{fig:mzagn}
\end{figure}

We are not able to assess AGN contamination for the full sample since we have $J$-band observations and significant detection of [OIII]$\lambda6584$ for only a fraction of our sample (72/162). In order to assess the potential impact of AGN contamination on our measurement of the MZ relation. We determine the MZ relation for the 87 galaxies where we have $J$ and $H$-band observations and compare this to the same sample with the 6 galaxies identified as AGN removed. Figure \ref{fig:mzagn} shows the MZ relation determined for the sample with (red points and curve) and without (black points and curves) the 6 galaxies identified as AGN removed. In this $J$-band observed subsample, AGN contamination leads to a slight overestimate of the metallicity for the least massive galaxies. However, the two relations are consistent within the observational errors and we conclude that AGN contamination is not significant in our sample. Throughout the rest of the paper, we remove the six galaxies identified as AGN. The results presented in Section 5 are based on analysis of 156 galaxies.

\subsection{Metallicity Comparison}

\begin{figure}
\begin{center}
\includegraphics[width=\columnwidth]{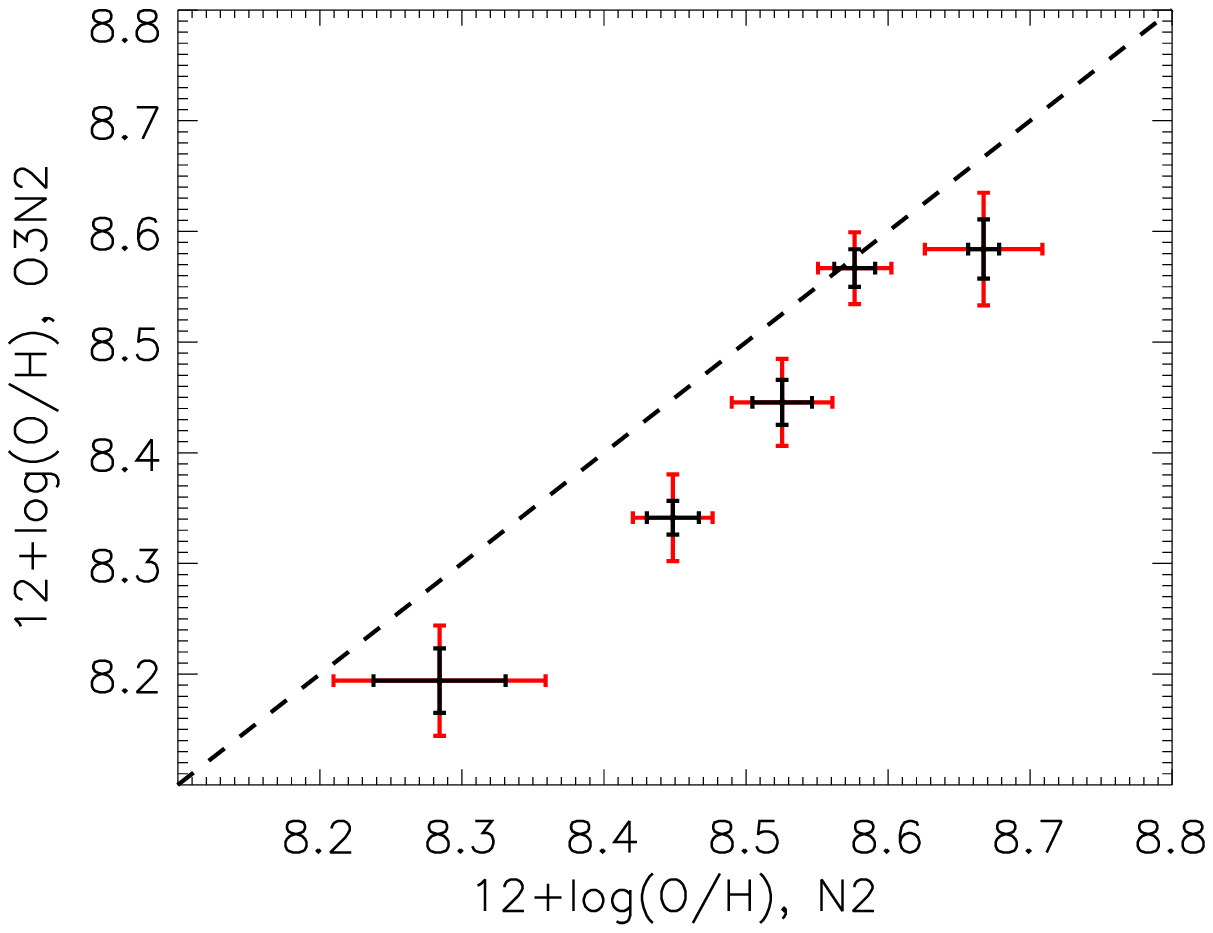}
\end{center}
\caption{A comparison of the metallicity determined from the $N2$ and $O3N2$ calibration of PP04 for the subsample of our galaxies with both $J$ and $H$-band observations. The metallicities are determined from spectra averaged in five mass bins. The black and red errors bars are the formal and bootstrapped errors, respectively. The dashed line is the one-to-one agreement.}
\label{fig:zcomp_pp04}
\end{figure}

In Figure \ref{fig:zcomp_pp04} we compare metallicities determined using the $N2$ and $O3N2$ diagnostics for the 81 galaxies for which we have both $J-$band and $H-$band spectroscopy. The data are sorted in the same manner as the previous section. The metallicities derived from the $O3N2$ diagnostic are systematically lower than those determined using the $N2$ diagnostic. This is result is consistent with \citet{Erb2006b} and \citet{Yabe2012} who also find that the $O3N2$ diagnostic gives systematically lower metallicities than the $N2$ diagnostic in high redshift galaxies. 

The $O3N2$ and the $N2$ diagnostic of \citet{Pettini2004} are calibrated to the same data and in the local Universe these two calibrations provide consistent results \citep[e.g.][]{Kewley2008}. However, Figure \ref{fig:zcomp_pp04} demonstrates that at higher redshifts, the two diagnostics are not consistent. The offset is even larger when using the more recent calibration of the $O3N2$ and $N2$ diagnostic provided by \citet{Marino2013}. The systematic offset in metallicity between the two diagnostics is attributed to changing physical conditions of the ISM in high redshift galaxies \citep{Erb2006b, Kewley2013b, Cullen2014}. In particular, a harder ionizing radiation field in high redshift galaxies is consistent with the $O3N2$ systematically offsetting to lower metallicities when compared with $N2$ (Kewley et al., In Preparation). In higher redshift galaxies, the high sensitivity of the $O3N2$ diagnostic to the ionization parameter makes it a poor indicator of metallicity (Kewley et al., In Preparation). The $N2$ diagnostic provides a more robust estimate and is the one adopted in this study. From our analysis we conclude that the $O3N2$ diagnostic should not be used to determine metallicities for galaxies outside the local Universe.

\section{Results}

\subsection{The Mass-Metallicity Relation}

\begin{deluxetable*}{ccccccc}
\tablecaption{FMOS-COSMOS MZ Relation Data}
\tablehead{\colhead{Stellar Mass} & \colhead{$N2$} &\colhead{12+log(O/H)} & \colhead{log($\Psi$)} & \colhead{$E(B-V)$}& \colhead{$N$} \\
 \colhead{log($M_\ast/M_\odot$)} &\colhead{log($[NII]/\mathrm{H}\alpha$)} & PP04 & $[M_\odot$ yr$^{-1}]$ & \colhead{Nebular}}
\startdata
9.70 & -0.951 $\pm$ 0.066 & 8.358 $\pm$ 0.038 & 1.175 $\pm$ 0.048 & 0.267 $\pm$ 0.012 & 16 \\
9.83 & -0.909 $\pm$ 0.060 & 8.382 $\pm$ 0.034 & 1.258 $\pm$ 0.059 & 0.279 $\pm$ 0.021 & 16 \\
9.95 & -0.783 $\pm$ 0.038 & 8.454 $\pm$ 0.022 & 1.397 $\pm$ 0.051 & 0.342 $\pm$ 0.020 & 16 \\
10.05 & -0.733 $\pm$ 0.038 & 8.482 $\pm$ 0.021 & 1.415 $\pm$ 0.056 & 0.357 $\pm$ 0.024 & 16 \\
10.11 & -0.678 $\pm$ 0.043 & 8.514 $\pm$ 0.024 & 1.509 $\pm$ 0.046 & 0.351 $\pm$ 0.034 & 16 \\
10.17 & -0.676 $\pm$ 0.034 & 8.515 $\pm$ 0.019 & 1.512 $\pm$ 0.076 & 0.403 $\pm$ 0.030 & 16 \\
10.25 & -0.647 $\pm$ 0.033 & 8.531 $\pm$ 0.019 & 1.702 $\pm$ 0.044 & 0.442 $\pm$ 0.032 & 15 \\
10.35 & -0.548 $\pm$ 0.024 & 8.587 $\pm$ 0.014 & 1.699 $\pm$ 0.054 & 0.466 $\pm$ 0.019 & 15 \\
10.50 & -0.435 $\pm$ 0.022 & 8.652 $\pm$ 0.013 & 1.824 $\pm$ 0.057 & 0.525 $\pm$ 0.037 & 15 \\
10.83 & -0.419 $\pm$ 0.023 & 8.661 $\pm$ 0.013 & 2.206 $\pm$ 0.093 & 0.679 $\pm$ 0.044 & 15 \\
\enddata
\label{tab:data}
\tablecomments{Column 1 gives the median stellar mass for each of the ten mass bins. Column 2 and 3 are the fitted $N2$ ratio and corresponding 12+log(O/H) using the PP04 calibration, respectively. {The errors only represent the observational uncertainties and do not include the 0.18 dex intrinsic dispersion of the metallicity calibration.} Column 4 and 5 are the median SFR and median $E(B-V)$ in each bin, respectively. The number of spectra averaged in each bin, $N$, is given in Column 6.}
\end{deluxetable*}

\begin{deluxetable}{ccccc}
\tablewidth{\columnwidth}
\tablecaption{MZ Relation Fit}
\tablehead{\colhead{Sample} & \colhead{Redshift} &\colhead{$Z_o$} & \colhead{$\mathrm{log}(M_o/M_\odot)$} & \colhead{$\gamma$}  }
\startdata
SDSS     & 0.08 & 8.710 $\pm$ 0.001 & 8.76 $\pm$ 0.01 & 0.66 $\pm$ 0.01  \\
COSMOS & 1.55 & 8.740 $\pm$ 0.042 & 9.93 $\pm$ 0.09 & 0.88 $\pm$ 0.18 \\

\enddata
\label{tab:fit}
\tablecomments{The fits shown in Figure \ref{fig:mz} and parameterized by Equation \ref{eq:fit}.}
\end{deluxetable}

\begin{figure}
\begin{center}
\includegraphics[width=\columnwidth]{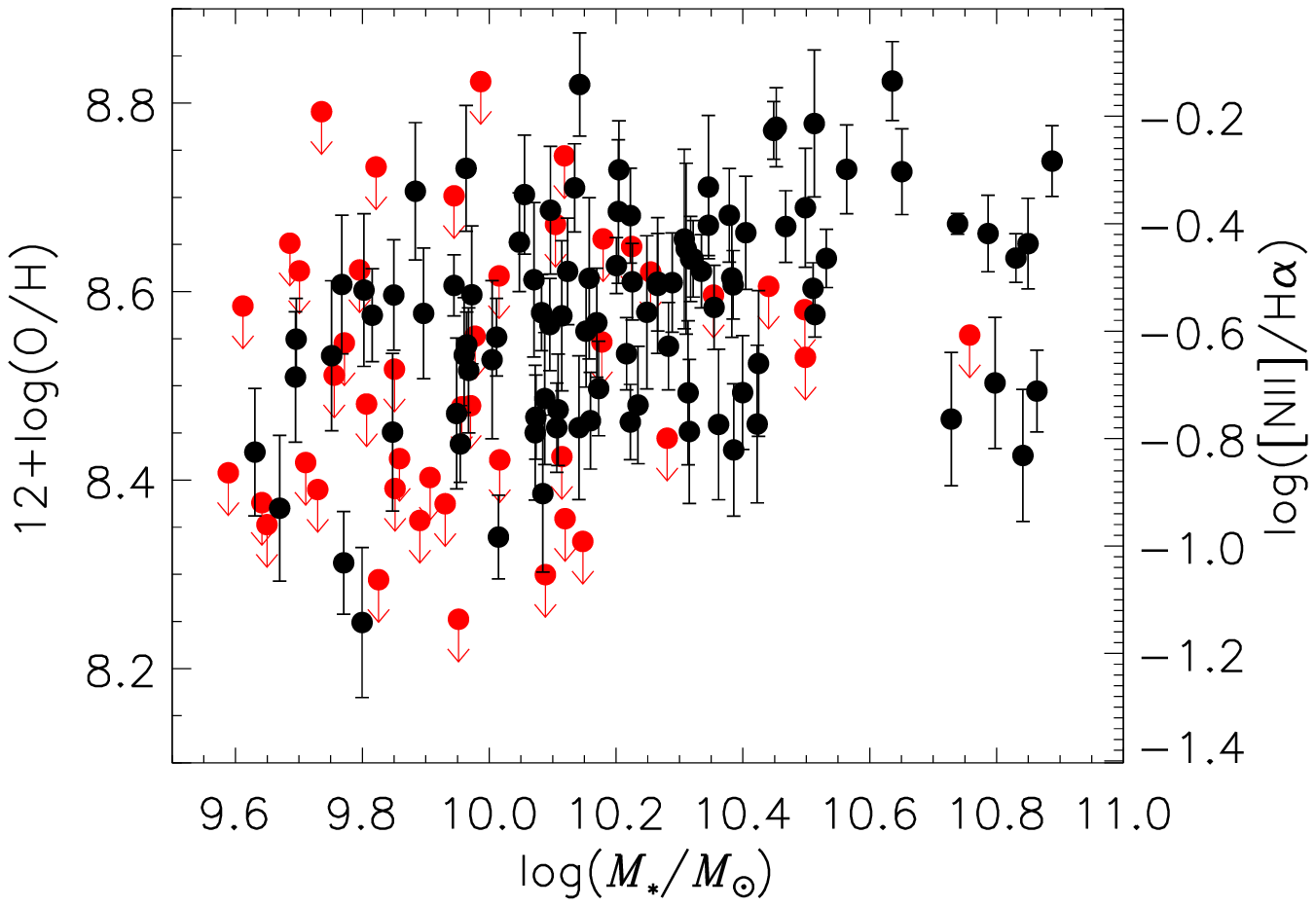}
\end{center}
\caption{The metallicity measured in individual galaxies as a function of stellar mass. The black data points are galaxies with [NII]$\lambda6584$ measured with S/N$>3$. The red points are galaxies for which we have adopted 3$\sigma$ upper limit for the [NII]$\lambda6584$ flux.  }
\label{fig:zgal}
\end{figure}

In Figure \ref{fig:zgal} we plot the metallicity as a function of stellar mass for individual galaxies in the sample. The metallicities are determined from the fitted $N2$ ratio using the PP04 calibration. The black points are the galaxies where [NII]$\lambda6584$ is measured with S/N$>3$. {The error bars only reflect the observational uncertainties and do not account for the 0.18 dex intrinsic dispersion of the metallicity calibration.} The red points are galaxies for which we have adopted an 3$\sigma$ upper limit for the [NII]$\lambda6584$ line flux. As Figure \ref{fig:zgal} demonstrates, we are more likely to not detect [NII]$\lambda6584$ in less massive galaxies. In Paper I we show that the less massive galaxies in our sample have lower SFRs. The greater number of non-detections in less massive galaxies is likely due to their lower SFRs. Strong sky lines contaminate $\sim10\%$ of the spectra and the small number of non-detections at higher stellar masses are likely due to strong sky line contamination.

Figure \ref{fig:zgal} shows that some less massive galaxies in our sample are metal-rich. The observed scatter in Figure \ref{fig:zgal} is the lower limit to the true scatter. Metal-rich galaxies are found across the whole stellar mass range probed in this study. This is consistent with the distribution  metal-rich galaxies observed in the local Universe \citep{Zahid2012a}. It is important to note that we have adopted a $3\sigma$ upper limit for [NII]$\lambda6584$ in galaxies. The number of galaxies with upper limits on metallicity is substantially larger at the lower mass end of the distribution. This suggests that the scatter likely increases as a function of stellar mass \citep[see also][]{Zahid2012a}. Here we have used the $N2$ diagnostic for determining metallicity and it is important to bear in mind that saturation of the diagnostic likely contributes in part to the nearly constant upper metallicity envelope of the distribution in Figure \ref{fig:zgal}.

Figure \ref{fig:zgal} demonstrates that our observations do not have sufficient sensitivity to provide an unbiased probe of metallicities in galaxies at stellar masses $M_\ast \lesssim 10^{10.3} M_\odot$. The S/N of [NII]$\lambda6584$ is a function of both SFR and metallicity. Since these physical properties are both strongly correlated with stellar mass, this leads to bias in the MZ relation determined from individual galaxies in our sample. We therefore rely on averaging spectra in order to increase the S/N of our measurement and determine a less biased MZ relation.

Our analysis is based on averaging spectra sorted into ten bins of stellar mass. There are 15 or 16 spectra in each of the mass bins. The stellar mass adopted is the median stellar mass in each bin. The metallicity is determined from the $N2$ line ratio determined from the averaged spectra shown in Figure \ref{fig:spectra}. The H$\alpha$ line is detected and SFRs are measured in individual objects without the necessity to stack (Paper I). The SFR adopted is the median SFR in each bin. The data are summarized in Table \ref{tab:data}.

\begin{figure*}
\begin{center}
\includegraphics[width=1.8\columnwidth]{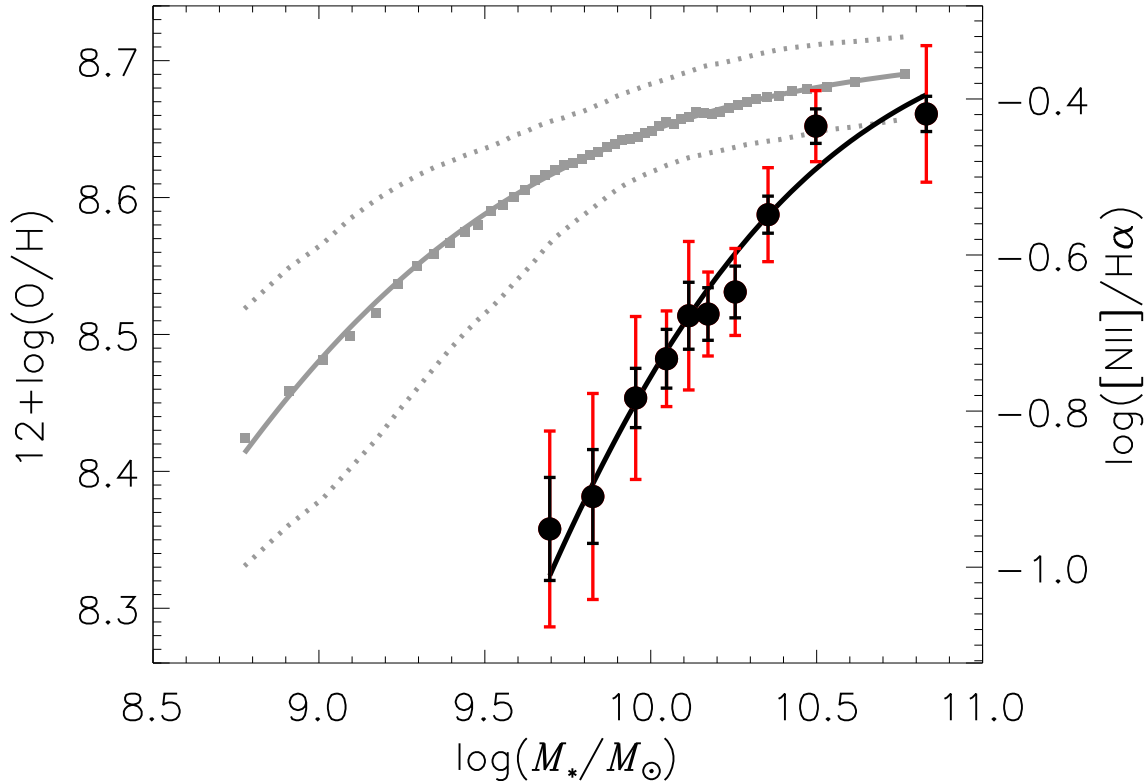}
\end{center}
\caption{The MZ relation determined from our sample of galaxies at $z\sim1.6$. The black points are metallicities determined from spectra averaged in ten mass bins. The black and red errors bars are the formal and bootstrapped errors, respectively. The black curve is a fit to the $z\sim1.6$ MZ relation as described by Equation \ref{eq:fit}. The gray data points are the median metallicities in 50 bins of stellar mass for galaxies in our local fiducial sample from SDSS. The solid gray curve is a fit to the local relation and the dotted lines denote the interval containing the central 68\% of galaxies.}
\label{fig:mz}
\end{figure*}

Figure \ref{fig:mz} shows the MZ relation at $z\sim1.6$ (black curve). The black points are the metallicities determined from the average spectra. The black error bars are the formal observational uncertainties determined by propagating the observational uncertainty in each pixel of the individual spectra. At a fixed stellar mass, there is intrinsic scatter in the metallicity distribution of galaxies \citep{Tremonti2004, Zahid2012a}. The red error bars are the standard deviation of the bootstrapped metallicity distribution in each bin of stellar mass and provide an estimate of the intrinsic scatter. Figure \ref{fig:mz} shows that in each of the stellar mass bins the observational uncertainties are always smaller than the intrinsic scatter.

We fit the MZ relation using the logarithmic form suggested by \citet{Moustakas2011}. The MZ relation is parameterized as
\begin{equation}
\mathrm{12 + log(O/H)} = Z_o - \mathrm{log}\left[1 + \left(\frac{M_\ast}{M_o}\right)^{- \gamma} \right].
\label{eq:fit}
\end{equation}
This function is preferred to a polynomial fit since it encapsulates much of our intuition regarding chemical evolution \citep[see][]{Moustakas2011, Zahid2013b}. In this parameterization, $Z_o$ is the asymptotic metallicity where the MZ relation flattens, $M_o$ is the characteristic stellar mass where the relation begins to flatten and $\gamma$ is the power-law slope of the MZ relation for stellar masses $<< M_o$. The parameters are determined using a $\chi^2$ minimization as implemented in the \emph{MPFIT} package in IDL \citep{Markwardt2009}. The MZ relation at $z\sim1.6$ is shown by the black curve. We derive the local MZ relation by sorting the data into 50 bins of stellar mass and taking the median metallicity in each bin. The error bars are smaller than the data points. The dotted curves contain the central 68\% of the galaxy distribution. We determine the MZ relation for the local sample using the same diagnostic (i.e. $N2$ calibrated by PP04) as applied to the $z\sim1.6$ data. The observational uncertainties are propagated through and the fit parameters and errors are given in Table \ref{tab:fit}.

In Section 4.1 and Figure \ref{fig:mzagn} we examine the MZ relation for our subsample for which we have $J$-band observations and are able to assess AGN contamination. The black curve in Figure \ref{fig:mzagn} is the MZ relation with six galaxies identified as AGN removed from the sample. The fit to the MZ relation in Figure \ref{fig:mzagn} is consistent with the relation we derive in Figure \ref{fig:mz}.

\subsection{The Stellar Mass, Metallicity and SFR Relation}

\begin{deluxetable*}{cccccc}
\tablecaption{FMOS-COSMOS Stellar Mass, Metallicity and SFR Relation}
\tablehead{\colhead{Stellar Mass} & \colhead{$N2$} &\colhead{12+log(O/H)} & \colhead{log($\Psi$)} & \colhead{$N$} \\
 \colhead{log($M_\ast/M_\odot$)} &\colhead{log($[NII]/\mathrm{H}\alpha$)} & PP04 & $[M_\odot$ yr$^{-1}]$}
\startdata
\hline
\noalign{\smallskip}
\multicolumn{5}{c}{Low-SFR} \\
\noalign{\smallskip}
\hline
\noalign{\smallskip}
9.82 & -0.803 $\pm$ 0.042 & 8.442 $\pm$ 0.024 & 1.155 $\pm$ 0.027 & 26 \\
10.11 & -0.660 $\pm$ 0.031 & 8.524 $\pm$ 0.018 & 1.415 $\pm$ 0.023 & 26 \\
10.36 & -0.534 $\pm$ 0.022 & 8.595 $\pm$ 0.013 & 1.614 $\pm$ 0.034 & 26 \\
\hline
\noalign{\smallskip}
\multicolumn{5}{c}{High-SFR}\\
\noalign{\smallskip}
\hline
\noalign{\smallskip}
9.89 & -0.916 $\pm$ 0.040 & 8.378 $\pm$ 0.023 & 1.397 $\pm$ 0.033 & 26 \\
10.16 & -0.683 $\pm$ 0.027 & 8.511 $\pm$ 0.015 & 1.665 $\pm$ 0.031 & 26 \\
10.56 & -0.477 $\pm$ 0.015 & 8.628 $\pm$ 0.009 & 1.999 $\pm$ 0.090 & 26 \\
\enddata
\label{tab:data_high_low}
\tablecomments{The same as Table \ref{tab:data} but split into two bins of SFR as described in the text.}
\end{deluxetable*}

The metallicities of galaxies are governed by gas flows and star formation. Recent work shows that in local galaxies there appears to be an \emph{anti-}correlation between the SFR and metallicity at a fixed stellar mass \citep{Ellison2008, Mannucci2010, Andrews2013}; at least at lower stellar masses \citep[see][]{Yates2012}. One possible explanation for this trend is that while inflows of gas dilute the gas-phase abundance and lower metallicities, they also fuel star-formation and lead to elevated SFRs. Motivated by this physical picture, \citet{Mannucci2010} propose that the parameterization of the relation between stellar mass, metallicity and SFR that minimizes the scatter in local galaxies is independent of redshift. They refer to this as the ``fundamental metallicity relation" (FMR). In their analysis, the lower metallicities of intermediate and high redshift galaxies are compensated by their higher SFRs such that galaxies out to $z\sim2.5$ are consistent with local FMR. However, it is clear that the FMR is highly dependent on methodology \citep{Yates2012, Andrews2013}. Therefore, applying a consistent methodology is important. We determine stellar masses, metallicities and SFRs for local galaxies and our sample at $z\sim1.6$ by applying as consistent a methodology as is currently possible in order to test the validity of the FMR.

The scatter in the local MZ relation is correlated with the SFR \citep{Ellison2008, Mannucci2010, Yates2012, Andrews2013}. We can examine the relation between stellar mass, metallicity and SFR for our sample at $z\sim1.6$ because we have SFRs determined from the H$\alpha$ luminosity in individual galaxies. We average the spectra by first sorting the data into three bins of stellar mass and then dividing the data in each bin of stellar mass into two bins of SFR. The metallicity and SFR in each bin of stellar mass and SFR is determined from 26 galaxies.

\begin{figure*}
\begin{center}
\includegraphics[width=1.8\columnwidth]{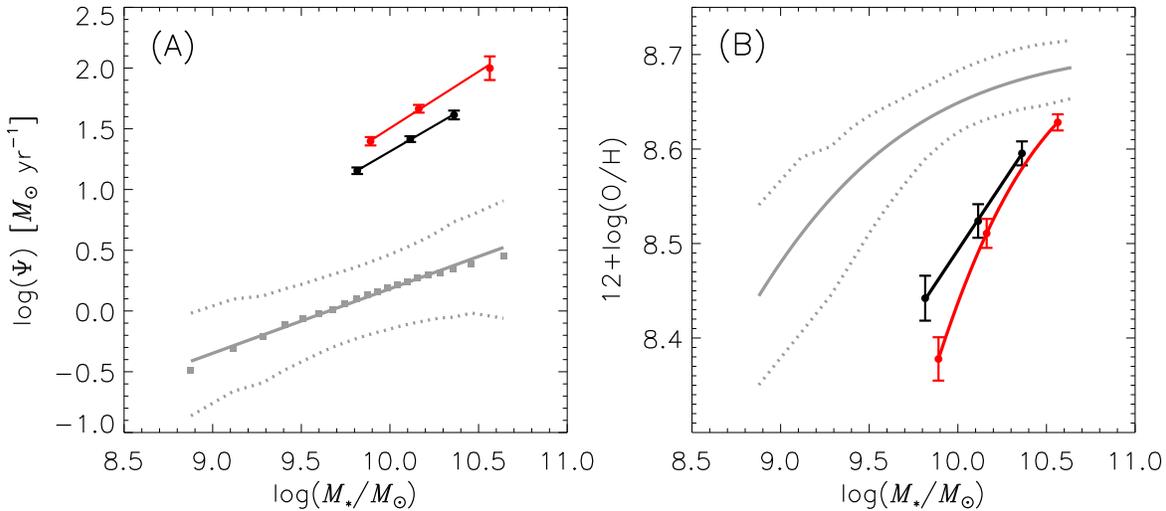}
\end{center}
\caption{(A) The relation between stellar mass and SFR. The black and red points are the median SFRs for galaxies that are first sorted into three mass bins and then two SFR bins. The solid black and red lines are linear fits to the relation between stellar mass and SFR for the high and low SFR bins, respectively. The gray points are the median SFRs sorted into 50 bins of stellar mass for star-forming galaxies in our local fiducial sample. The solid line is a fit to the relation between stellar mass and SFR and the dotted line is denotes the interval containing central 68\% of galaxies. (B) The MZ relation for the sample divided into bins of stellar mass and SFR for the same data shown in (A). The black and red points are the metallicities in bins of stellar mass and SFR. The black and red curves are fits to the high and low SFR data, respectively. The gray curve is the fit to the local MZ relation and the dotted line denotes the interval containing the central 68\% of galaxies.}
\label{fig:mzsfr}
\end{figure*}

In Figure \ref{fig:mzsfr}A we plot the median SFR in each bin. The error bars are the bootstrapped errors on the median and are analogous to the standard error on the mean. The red and black curves in Figure \ref{fig:mzsfr} are linear fits with a slope of $0.93\pm0.13$ and $0.85\pm0.08$, respectively. The local stellar mass - SFR relation is shown in gray. The local relation is determined by taking the median SFR in 50 bins of stellar mass. The solid gray line is a fit to the local relation. The dotted gray curves indicate the limits containing the central 68\% of the galaxy distribution. The slope of the local relation is $0.68\pm0.01$ and does not differ significantly ($2\sigma$) from the slope of the $z\sim1.6$ relation. However, the slope of the relation is dependent on sample selection \citep[cf.][]{Peng2010}.

In Figure \ref{fig:mzsfr}B we show the MZ relation for the data sorted into bins of SFR. The MZ relation shown by the red and black curve correspond to the high and low SFR bins, respectively, shown in Figure \ref{fig:mzsfr}A. The gray curve is the local MZ relation. The metallicities of high SFR galaxies (red curve) are systematically lower than the metallicities of low SFR galaxies (black curve). At a fixed stellar mass, the metallicity is \emph{anti}-correlated to the SFR. This is similar to trends seen in local galaxies at lower stellar masses \citep[e.g.][]{Mannucci2010}. It is important to note that while the slope of the stellar mass - SFR relation is similar for local and $z\sim1.6$ galaxies, the MZ relation for the same galaxies is significantly steeper. This has important implications for the FMR proposed by \citep{Mannucci2010}. We examine this issue in detail in Section 7.3.

\subsection{The Stellar Mass, Metallicity and E(B-V) Relation}

\begin{deluxetable*}{cccccc}
\tablecaption{FMOS-COSMOS Stellar Mass, Metallicity and E(B-V) Relation}
\tablehead{\colhead{Stellar Mass} & \colhead{$N2$} &\colhead{12+log(O/H)} & \colhead{E(B-V)} & \colhead{$N$} \\
 \colhead{log($M_\ast/M_\odot$)} &\colhead{log($[NII]/\mathrm{H}\alpha$)} & PP04 & Nebular}
\startdata
\hline
\noalign{\smallskip}
\multicolumn{5}{c}{Low-E(B-V)} \\
\noalign{\smallskip}
\hline
\noalign{\smallskip}
9.77 & -0.963 $\pm$ 0.046 & 8.351 $\pm$ 0.026 & 0.259 $\pm$ 0.008 & 26 \\
10.11 & -0.734 $\pm$ 0.027 & 8.482 $\pm$ 0.015 & 0.325 $\pm$ 0.012 & 26 \\
10.38 & -0.564 $\pm$ 0.019 & 8.579 $\pm$ 0.011 & 0.464 $\pm$ 0.015 & 26 \\
\hline
\noalign{\smallskip}
\multicolumn{5}{c}{High-E(B-V)}\\
\noalign{\smallskip}
\hline
\noalign{\smallskip}
9.95 & -0.782 $\pm$ 0.036 & 8.454 $\pm$ 0.020 & 0.342 $\pm$ 0.012 & 26 \\
10.16 & -0.622 $\pm$ 0.031 & 8.545 $\pm$ 0.018 & 0.446 $\pm$ 0.008 & 26 \\
10.64 & -0.395 $\pm$ 0.018 & 8.675 $\pm$ 0.010 & 0.637 $\pm$ 0.024 & 26 \\
\enddata
\label{tab:data_high_low}
\tablecomments{The same as Table \ref{tab:data} but split into two bins of $E(B-V)$ as described in the text.}
\end{deluxetable*}

\begin{figure*}
\begin{center}
\includegraphics[width=1.8\columnwidth]{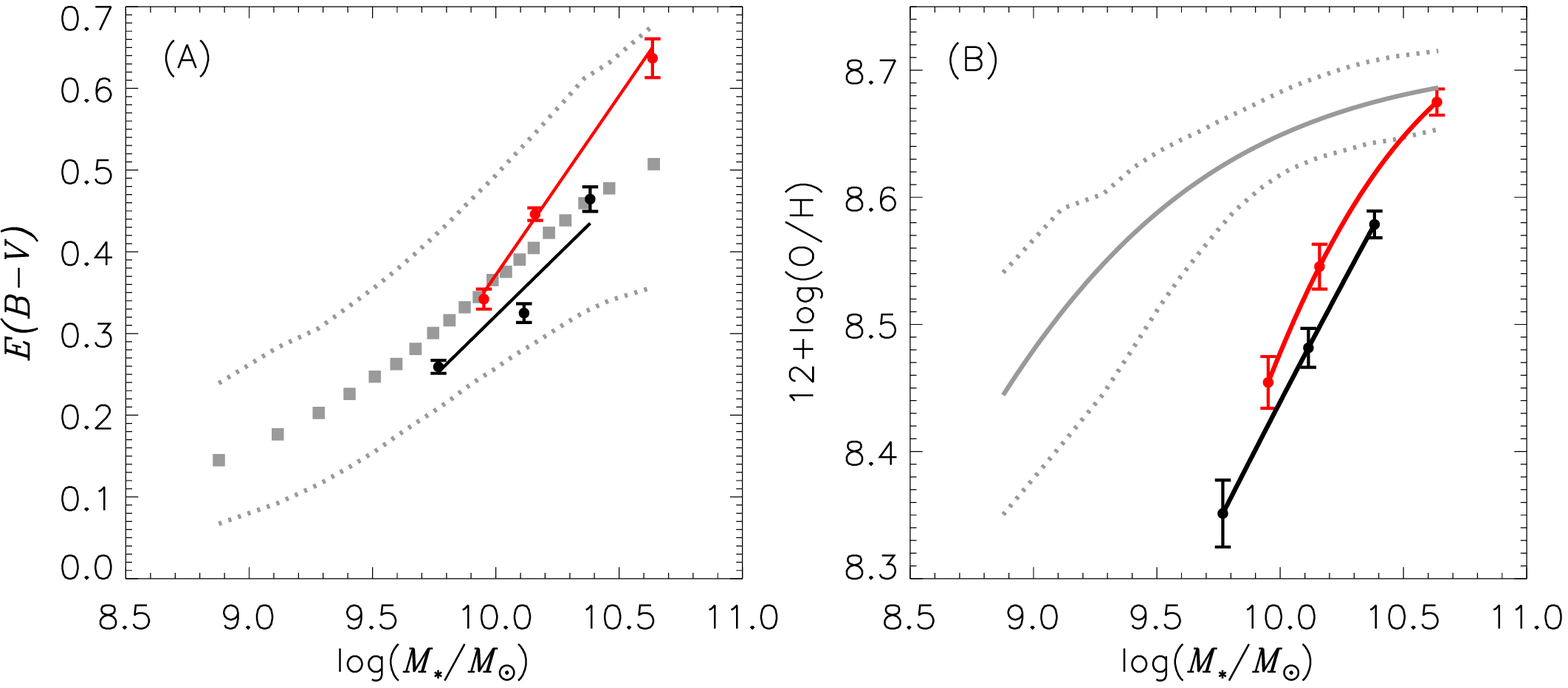}
\end{center}
\caption{(A) The relation between stellar mass and $E(B-V)$. The black and red points are the median nebular $E(B-V)$ for galaxies that are first sorted into three mass bins and then two $E(B-V)$ bins. The solid black and red lines are linear fits to the relation between stellar mass and $E(B-V)$ for the high and low $E(B-V)$ bins, respectively. The gray points are the median $E(B-V)$ sorted into 50 bins of stellar mass for star-forming galaxies in our local fiducial sample. The solid line is a fit to the relation between stellar mass and SFR and the dotted lines denote the interval containing central 68\% of galaxies. (B) The MZ relation for the sample divided into bins of stellar mass and $E(B-V)$ for the same data shown in (A). The black and red points are the metallicities in bins of stellar mass and $E(B-V)$. The black and red curves are fits to the high and low SFR data, respectively. The gray curve is the fit to the local MZ relation and the dotted line denotes the interval containing the central 68\% of galaxies.}
\label{fig:ebv}
\end{figure*}

A strong correlation is observed between metallicity and dust in the local Universe \citep{Heckman1998, Boissier2004, Asari2007, Garn2010b, Xiao2012, Zahid2012b, Zahid2013a}. \citet{Reddy2010} indirectly examine the relation between dust obscuration and metallicity at $z\sim2$. They find that the $L_{IR}/L_{UV}$ ratio, which is taken as a proxy for dust, scales with stellar mass. They combine the relation between stellar mass and $L_{IR}/L_{UV}$ ratio with the relation between stellar mass and metallicity from \citet{Erb2006b} to derive a relation between metallicity and $L_{IR}/L_{UV}$ ratio, i.e. dust extinction.

In Figure \ref{fig:ebv} we examine the relation between stellar mass, metallicity and $E(B-V)$. We average the spectra by first sorting the data into three bins of stellar mass and then divide the data in each bin of stellar mass into two bins of $E(B-V)$. The metallicity and $E(B-V)$ in each bin is determined from 26 galaxies. In Figure \ref{fig:ebv}A the $E(B-V)$ value is the median of the 26 galaxies in each bin and the errors are bootstrapped. The red and black points correspond to the high and low $E(B-V)$ bins, respectively. The gray squares are the median $E(B-V)$ values in 50 bins of stellar mass for the local sample of galaxies. The dotted lines denotes the interval containing 68\% of the galaxy distribution. 

In Figure \ref{fig:ebv}B we plot the metallicity determined in bins of stellar mass and $E(B-V)$. The red and black points and curves correspond to the high and low $E(B-V)$ sample, respectively. Figure \ref{fig:ebv}B shows that galaxies with higher $E(B-V)$ (red points) also have higher metallicity. At a fixed stellar mass, the metallicity is correlated with dust extinction. Similar trends are observed in the local sample \citep[e.g.,][]{Zahid2012b, Zahid2013a}. This correlation is perhaps not that surprising since dust is composed of metals.

\section{Systematic Issues in the Measurements}

Before turning to a discussion of our results, we highlight some of the systematic issues in our measurements. Current and future spectroscopic surveys should be able to address many of these issues. 

The fit to the MZ relation presented in Section 5.1 is strongly constrained by the two highest stellar mass bins. In Figure \ref{fig:mz}, the bin at $M_\ast \sim 10^{10.5} M_\odot$ is above the relation. Figure \ref{fig:zgal} shows that there is a lack of intermediate and low metallicity objects for galaxies in this mass range. If we exclude the bin at $M_\ast \sim 10^{10.5} M_\odot$, the fitted $Z_o$ and $\mathrm{log}(M_o/M_\odot)$ are $8.75 \pm 0.06$ and $9.94 \pm 0.15$, respectively. These values are consistent with values given in Table \ref{tab:fit}. Excluding the second to highest stellar mass bin in our fit does not change any of the conclusions of this paper. However, if we exclude the highest stellar mass bin the two fits are not consistent. The saturation metallicity, $Z_o$, and turnover mass, $M_o$, are both strongly dependent on the highest mass bin. 

{Our analysis of AGN contamination presented in Figure \ref{fig:bpt} suggests that AGN contamination does not significantly affect our measurement of the MZ relation. Similarly, \citet{Wuyts2014} find that MZ relations they measure at $z\sim1$ and $z\sim2$ are insensitive to AGN contamination. This might be due to evolving ISM conditions which may result in a BPT diagram where the [NII]/H$\alpha$ line ratios of AGN overlap with star-forming galaxies \citep[e.g.,][]{Kewley2013a, Kewley2013b}. Whatever the case, Figure 4 and results from \citet{Wuyts2014} suggest that AGN contamination is not a significant source of bias.} 

Our sample presented here is selected primarily using the sBzK selection in order to maximize detections. 
{In Paper III we show that the an sBzK sample spans the same color space as a sample selected on the basis of full SEDs. There does not seem to be a serious bias with the sBzK sample that would affect the metallicity.  In Paper III we show that the sBzK selection produces a narrower main sequence. This may decrease some of the dispersion in the intrinsic properties of the sample but should not cause any systematic offset that could change the slope of the MZ relation. }

While the sBzK selection is not significantly biased against dusty galaxies because the reddening vector moves objects parallel to the selection cut \citep{Daddi2004}, our effective sensitivity of $4\times10^{-17}$ ergs s$^{-1}$ cm$^{-2}$ may not reach deep enough to observe H$\alpha$ in objects that are heavily obscured by dust. To increase likelihood of detection, for the majority of the sample we imposed a second selection criteria that required obscured SFRs of $\sim5 M_\odot$ yr$^{-1}$ based on UV luminosities. This introduces bias against heavily obscured objects. In Figure \ref{fig:ebv} we show that at the metallicity is correlated with dust extinction. Thus, if our sample is biased against heavily obscured objects, we may be missing many metal-rich galaxies in our sample $z\sim1.6$. This bias is likely to be mass dependent and effect our measurement of the MZ relation at high stellar masses. {A correction for this bias would not lower the slope of the MZ relation.} However, galaxies with heavy dust obscuration also tend to have higher SFRs. The relation between SFR and metallicity is opposite to the relation between dust extinction and metallicity and therefore these two effects may offset each other. The relation between stellar mass, metallicity, dust extinction and SFR are complicated and larger spectroscopic samples are required in order to assess any bias that may be present in our measurements. 

{The observed anti-correlation between metallicity and SFR \citep[e.g.][]{Mannucci2010} may lead to an overestimate of the MZ relation at low stellar masses. This is because the galaxies with SFRs falling below our detection limit are preferentially found at lower stellar masses and because of the anti-correlation between metallicity and SFR, these low SFR galaxies also tend to have high metallicities. This type of bias would effect the lowest stellar mass bins resulting in an artificial steepening of the MZ relation. However, the SFRs we measure show the same amount of scatter at low and high stellar masses \citep[see][]{Kashino2013} and our $K$-band selection limit is designed to select galaxies with stellar masses where the SFRs are detectable. While the distribution of SFRs we measure is slightly narrower than previous studies \citep[see][for discussion]{Kashino2013}, the survey is designed such that we are not missing a significant number of galaxies with low SFRs. Thus, this type of bias is unlikely to significantly change the slope we measure.}

One of the primary challenges with spectroscopically accessing the redshift desert ($1 \lesssim z \lesssim 2$) is that optical emission lines are redshifted into the near-infrared. The near-infrared is significantly contaminated by strong atmospheric emission lines, making observations of faint emission lines extremely difficult. This is compounded by detector throughput and sensitivities that are significantly below optical detectors. Because of these difficulties, we are not able to observe faint emission lines such as [NII]$\lambda6584$ in a significant number of individual galaxies. Instead, we must rely on averaging many spectra in order to achieve the S/N necessary to detect weak emission lines. This may be problematic since the properties of galaxies that we are measuring are not necessarily linear with line strength. For example, the [NII]$\lambda6584$ line strength scales exponentially with metallicity and therefore simply averaging spectra may bias our measurement. 

{To address the issue of stacking, we have measured the relation using the median of the stacked spectra, rather than average to test for any systematic bias. The two methods yield consistent results. Additionally, we examine the bias in local star-forming galaxies by sorting galaxies into 50 bins of stellar mass and then averaging the line flux of [NII]$\lambda6584$ and H$\alpha$ \emph{before} determining metallicity. Thus, we determine the MZ relation from the average [NII]$\lambda6584$ and H$\alpha$ flux. We find a very small offset ($\sim 0.01$ dex) between the two methods. In \citet{Geller2014}, we measure the MZ relation determined from stacked spectra using the R23 diagnostic and compare with the MZ relation we derive in \citet{Zahid2013b} from individual galaxies (see Figure 16 of Geller et al.). The  stacked data give results consistent with measurements of the MZ relation derived from individual galaxies. We conclude that the MZ relation derived from stacked data is consistent with the MZ relation derived from individual galaxies.} 

{For many objects, the redshift is determined from a single emission line. In Paper III we compare our redshifts with the zCOSMOS spectroscopic redshift survey \citep{Lilly2007}. There are 37 objects in both catalogs which allow us to assess the accuracy of our redshift determination. We find that 33/37 objects ($89\%$) yield consistent redshifts. Some fraction of the inconsistent redshifts are due to errors in the zCOSMOS assignment. We conclude that misidentification affects $\lesssim10\%$ of the sample. Misidentification of an emission line results in an underestimate of the [NII]$\lambda6584$ flux relative to H$\alpha$ since there is no corresponding [NII]$\lambda6584$ emission. Assuming that [NII]$\lambda6584$ fluxes are distributed about the mean, we estimate that misidentification fraction translates directly into the fractional underestimate of the average flux measured in our stacked data. A $\lesssim10\%$ error in [NII]$\lambda6584$ flux translates to a $\lesssim0.03$ dex underestimate of the uncertainty. This level of contamination does not change any of the major conclusions of this work.}

We have used the local calibrations for metallicity and applied them to high redshift data. Several authors have shown that evolving ISM conditions may lead to evolution in key emission line diagnostics \citep{Erb2006b, Hainline2009, Rigby2011, Yabe2012, Kewley2013a}. Typically, these studies have argued for evolving ISM conditions on the basis of the [OIII]/H$\beta$ vs. [NII]/H$\alpha$ diagram. However, the metallicities of galaxies are also dependent on conditions of the ISM \citep[e.g.][]{Kewley2002a}. In this study, we have shown that the metallicities determined using the $N2$ and $O3N2$ line ratios are inconsistent despite being calibrated to the same data in the local Universe (PP04). In order to assess the impact of these variations on metallicity, deep observations of full optical spectra in a large sample of galaxies are necessary.

In \citet{Zahid2013b} we attribute the flattening of the MZ relation for massive galaxies to the physical effect of metallicity saturation. In that study, we determine metallicities using the $R23$ diagnostic. Photoionization modeling suggests that this metallicity diagnostic saturates at metallicities significantly higher than the maximum metallicity observed in star-forming galaxies \citep{Kewley2002a}. However, the $N2$ diagnostic is prone to saturation at significantly lower metallicities \citep{Kewley2002a}. The saturation of $N2$ suggests that metallicities may be higher than those that we have measured here. However, we emphasize that the flattening observed in the local MZ relation is present using several different diagnostics \citep[see][]{Kewley2008}. Moreover, the high $N2$ ratios are only observed in the most massive galaxies and therefore saturation may only effect a small fraction of the most massive galaxies. Since we have applied the same diagnostic to both or local and $z\sim1.6$ sample, we have mitigated uncertainties in the relative metallicities. 

In Figure \ref{fig:mzsfr} we examine the relation between stellar mass, metallicity and SFR. Similar to the local Universe, there appears to be an anti-correlation between metallicity and SFR at a fixed stellar mass. We measure SFR from the H$\alpha$ line flux and metallicity is determined from the $N2$ ratio. While SFR is directly dependent on the H$\alpha$ line flux, the metallicity is inversely dependent on the H$\alpha$ line flux. The errors are correlated in the same way as the observed trends. This could potentially produce an artificial trend in our measurement of the stellar mass, metallicity and SFR relation for galaxies at $z\sim1.6$. However, we note that for local galaxies such trends persist even when completely independent line diagnostics are used. For example, similar trends are observed when SFRs are determined from Balmer lines and metallicities are determined from the [NII]$\lambda6584$/[OII]$\lambda3727$ ratio \citep{Andrews2013}. In this case, the two measurements are independent and the effect in local galaxies can not be attributed to correlated errors. It remains to be seen whether trends persist when using independent diagnostics for higher redshift samples.

In Figure \ref{fig:ebv} we examine the relation between stellar mass, dust extinction and metallicity. The trends observed in the relation at $z\sim1.6$ are similar to those observed in local galaxies. Namely, at a fixed stellar mass, dust extinction is correlated to metallicity. However, we are currently not able to apply a consistent methodology when examining these two samples because H$\beta$ is detected in only a small fraction of our sample. We instead rely on dust extinction determined from the continuum extinction measured from the $B-z$ color. We convert this to nebular extinction using the factor we derive in Paper I. While this may apply to population on average, this may not be applicable to individual galaxies. Future surveys with higher sensitivity and broader wavelength coverage should be able to establish the validity of this approach and robustly establish the relation between stellar mass, dust extinction and metallicity.

\section{Discussion}

\subsection{Comparison of the MZ Relation with Previous High-z Studies}

\citet{Yabe2012} report on the MZR at $z\sim1.4$ based on FMOS observations conducted in low resolution mode. The initial sample is $K$-band selected with a secondary selection for galaxies expected to have H$\alpha$ flux $>10^{-16}$ ergs s$^{-1}$ cm$^{-2}$ based on rest-frame UV emission. They derive an MZR by stacking the spectra of 71 galaxies that have significant H$\alpha$ detections in 3 mass bins. In order to make a robust comparison, both the stellar masses and metallicities must be determined in a consistent manner. \citet{Yabe2012} derive metallicities from the $N2$ using the PP04 calibration. However, we find that the stellar mass estimates are systematically offset mostly (but not completely) due to the different IMFs adopted. We recalculate the stellar masses applying our methodology using photometry provided by K. Yabe. We compare our derived stellar masses with those calculated by \citet{Yabe2012}. Our mass estimates are systematically lower by 0.28 dex. For consistency, we subtract 0.28 dex from the stellar masses derived by \citet{Yabe2012} when plotting the MZR.

\begin{figure}
\begin{center}
\includegraphics[width=\columnwidth]{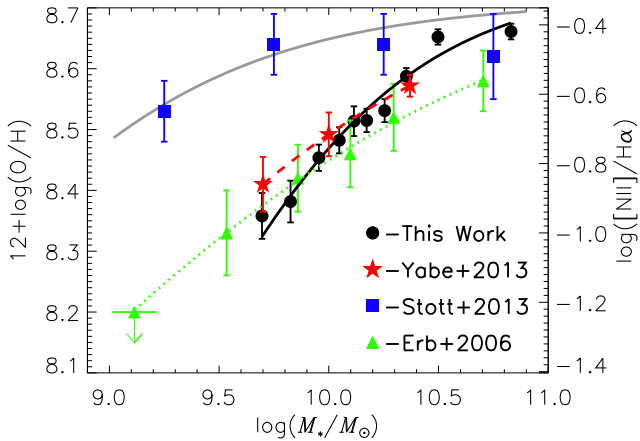}
\end{center}
\caption{A comparison of the MZ relation we measure at $z\sim1.6$ with the measurement at $z\sim1.4$ from \citet{Yabe2012}. The black filled circles and curve are our measurements. The red stars are measurements from \citet{Yabe2012}.}. 
\label{fig:yabe}
\end{figure}

Figure \ref{fig:yabe} shows a direct comparison between the MZR derived by \citet[red stars]{Yabe2012} and the MZR we measure (black solid curve and filled circles). Our data extend $\sim0.5$ dex higher in stellar mass. The MZR we derive is systematically steeper, though over the stellar mass range probed by \citet{Yabe2012} the metallicities in the individual bins are consistent within the errors. As we noted in Section 6, some small fraction of the galaxies have misidentified H$\alpha$. Including these galaxies in the average leads to an underestimate of the metallicity. The misidentification of H$\alpha$ largely effects the least massive galaxies in the sample. This may explain the lower metallicities we measure in the lowest mass bin. Given the differences in sample size (156 compared to 71 galaxies) and observational modes (high resolution compared to low resolution) we consider the good agreement in the two measurements to be reassuring. 

In Figure \ref{fig:yabe} we also plot the MZ relation from \citet[green triangles]{Erb2006b}. \citet{Erb2006b} average the spectra of 87 star-forming galaxies at $z\sim2.3$ in five bins of stellar mass. We have recalculated their stellar masses using the same methodology applied to our $z\sim1.6$ sample in order to ensure a consistent comparison \citep[see][]{Zahid2012b}. The \citet{Erb2006b} sample is flatter than the relation we derive for our sample at $z\sim1.6$. At the low mass end, our metallicities may be underestimated. This may explain part of the discrepancy (see Section 6). At the high mass end, it is possible that metallicities determined by \citet{Erb2006b} may be underestimated. The sample of \citet{Erb2006b} is UV selected Lyman break galaxies and therefore biased against dusty objects. In Figure \ref{fig:ebv} we show that dust extinction and metallicity are correlated. The UV selected samples are likely to be missing the dustier, metal-rich massive galaxies and therefore the average metallicity derived by \citet{Erb2006b} may be underestimated in the highest mass bins.

\citet{Henry2013} have examined the MZ relation for low mass galaxies at $z\sim1.8$. The data demonstrate a clear decline in metallicity at lower stellar masses (down to $\sim10^{8}M_\odot$). The relation they derive is consistent with our measurements. However, we note that due the faintness of low mass galaxies, the observational uncertainties are large and therefore do not provide a strong validation of the MZ relation we derive.

Recently, \citet{Stott2013} report a MZ relation for a combined sample of galaxies at $z=0.84$ and $z=1.47$. They determine the metallicity from stacking spectra 103 galaxies into four bins of stellar mass. The blue squares in Figure \ref{fig:yabe} are their measurements of the MZ relation. The primary conclusion of \citet{Stott2013} is that the MZ relation does not evolve with redshift (compare their data with local relation shown by the gray curve in Figure \ref{fig:yabe}). They argue that the more than dozen previous studies reporting an evolution in the MZ relation are biased. They cite the higher SFRs probed in previous studies and selection bias in UV selected samples as the origin of the reported evolution. 

Figure \ref{fig:yabe} clearly demonstrates that the lack of evolution in the MZ relation reported by \citet{Stott2013} is not supported by our data. The potential sources of bias provided by \citet{Stott2013} do not strictly apply to our data. Our sample is based on the sBzK selection which is significantly less biased against dusty objects as compared to UV selections. This is because in color-color space, the effect of dust is to move objects parallel to the selection criteria \citep{Daddi2004}. Furthermore, this effect should be most pronounced for massive galaxies and low mass galaxies are unlikely to be severely dust obscured. A second source of bias suggested by \citet{Stott2013} is that previous studies probe significantly higher average SFRs. We note that our observations are carried using FMOS on Subaru operated in high resolution mode. This setup is identical to the observational setup used by \citet{Stott2013}. However, our observations are significantly deeper as we observe a single pointing position per night whereas \citet{Stott2013} observed six positions in a night. Our observations have a sensitivity limit that is nearly an order of magnitude deeper than the observations of \citet{Stott2013} and therefore we are able to to observe galaxies with significantly lower SFRs. We conclude that our sample does not suffer significantly from the type of bias suggested by \citet{Stott2013}. We consider the redshift evolution of the MZ relation to be real.

\subsection{Evolution of the MZ relation}


In \citet{Zahid2013b} we examine the evolution of the MZ relation. Our analysis is primarily based on three large samples at $z<1$ for which we are able to measure metallicities in individual galaxies using the same metallicity calibration \citep[i.e.][]{Kobulnicky2004}. In \citet{Zahid2013b}, we fit the MZ relation using the parameterization given in Equation \ref{eq:fit} and conclude that the shape of the MZ relation evolves significantly with redshift such that it flattens at late times. In an ``open-box" model of galaxy chemical evolution where star formation is primarily fueled by cosmological accretion and is capable of driving large scale galaxy winds which expel metals from the ISM, the gas-phase oxygen abundance may not exceed the effective yield.\footnote{The nucleosynthetic yield is the mass of oxygen formed per unit SFR. In the presence of outflows, some metals may be lost and the effective yield is the mass of oxygen produced minus the mass of oxygen lost in the wind per unit SFR.} We argue that the flattening of the MZ relation is primarily driven by massive galaxies enriching to the effective yield. We show that  upper metallicity limit, $Z_o$, does not to evolve significantly out to \emph{at least} $z\sim0.8$ \citep{Zahid2013b}. Furthermore, we show that the flattening of the relation can be primarily understood as an evolution in $M_o$, the stellar mass at which the MZ relation begins to flatten. $M_o$ is $\sim0.7$ dex lower in the local universe as compared to $z\sim0.8$ \citep{Zahid2013b}.

The MZ relation we derive at $z\sim1.6$ is consistent with the evolution observed in the MZ relation at $z<1$. Table \ref{tab:fit} gives the fit parameters for the local and $z\sim1.6$ relation\footnote{The fit parameters given in Table \ref{tab:fit} can not be directly compared to those provided in \citet{Zahid2013b}. Because of the different spectral ranges covered by the data, we are required to apply different metallicity calibrations in determining metallicities. Systematic differences between various metallicity calibrations are well documented, though relative metallicities are found to be robust \citep[e.g.][]{Kewley2008}}. Within the observational uncertainties, the saturation metallicity, $Z_o$, does not evolve since $z<1.6$ and $M_o$ is $\sim1.2$ dex larger at $z\sim1.6$. The lack of evolution in the saturation metallicity places constraints on the metallicity of outflows and oxygen yields in star-forming galaxies. Detailed comparison of the evolution with analytical and numerical models is needed to rigorously establish this. The lack of evolution in $Z_o$ and the evolution of $M_o$ to larger masses means that only the most massive galaxies at $z\sim1.6$ achieve the level of enrichment observed in the local universe. However, our interpretation remains tentative due to possible saturation of the $N2$ diagnostic (see Section 6). 

The high gas-phase abundance of massive star-forming galaxies at $z\sim1.6$ is consistent with observations which indicate super-solar metallicities for the stellar populations of massive early-type galaxies \citep[e.g.][]{Gallazzi2005, Thomas2005, Panter2008, Conroy2013a}. Analysis of the stellar population ages of these massive early type galaxies indicates old stellar populations ($\gtrsim10$ Gyr). This implies that these galaxies formed stars in the distant past from gas enriched to the level observed in massive star-forming galaxies in the local universe. More to the point, \citet{Panter2008} show that the most massive early-type galaxies exhibit super-solar stellar metallicities that do not evolve out to $z\gtrsim2$. In contrast, less massive galaxies show significant evolution since $z\sim2$ with the least massive galaxies showing the greatest evolution. The stellar metallicity evolution trends observed by \citet{Panter2008} are consistent with the evolution observed in the gas-phase oxygen abundance of the star-forming galaxy population since $z\sim1.6$, i.e. a steepening of the MZ relation at higher redshift. \emph{The chemical evolution of star-forming galaxies may be characterized by an upper metallicity limit that does not evolve with redshift. This upper limit is achieved in progressively lower stellar mass galaxies as the Universe evolves.} This evolution is likely driven by the decreasing gas fraction of galaxies with time.

\subsection{The Stellar Mass, Metallicity and SFR Relation}

\begin{figure}[t!]
\begin{center}
\includegraphics[width=\columnwidth]{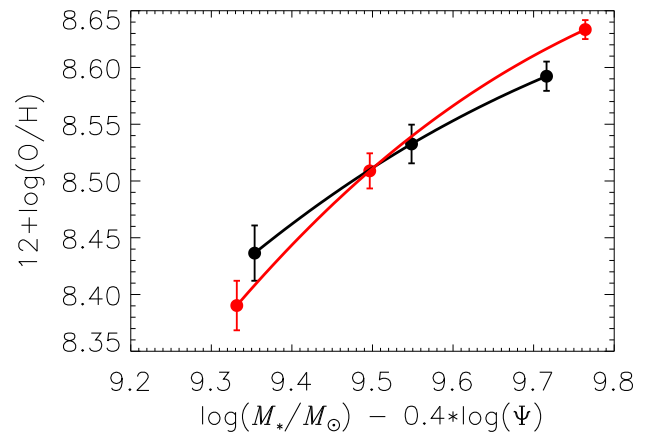}
\end{center}
\caption{Following the approach of \citet{Mannucci2010}, we plot metallicity against the combination of stellar mass and SFR that minimizes the scatter for our sample at $z\sim1.6$. The data are the same as in Figure \ref{fig:mzsfr}.}
\label{fig:fmr1}
\end{figure}

\citet{Ellison2008} first showed the \emph{anti}-correlation between specific SFR and metallicity at a fixed stellar masses. Subsequently, this relation has been examined by several authors \citep[e.g.][and others]{Mannucci2010, Yates2012, Andrews2013}. These studies show that for SDSS galaxies, the scatter around the MZ relation is somehow correlated to the SFR. In particular, \citet{Mannucci2010} find that at a fixed stellar mass, the metallicity is \emph{anti-}correlated to the SFR. They parameterize the metallicity as a function of stellar mass and SFR, i.e. $\mu_\alpha = \mathrm{log}(M_\ast/M_\odot) - \alpha\mathrm{log}(\Psi)$. They find that $\alpha=0.32$ minimizes the scatter in the local MZR. They argue that the relation between metallicity and $\mu_\alpha$ does not evolve for galaxies with $z<3$ and refer to this relation as the ``fundamental metallicity relation" (FMR). For the FMR, the lower metallicities observed in high redshift galaxies are compensated by their higher SFRs. However, the relation is dependent on methodology \citep{Yates2012, Andrews2013}. There is no consensus regarding the evolution of the relation between stellar mass, metallicity and SFR and therefore the validity of the FMR remains tentative \citep[e.g][]{Niino2012, Perez-Montero2013, Sanchez2013, Cullen2014}.

\begin{figure}[t!]
\begin{center}
\includegraphics[width=\columnwidth]{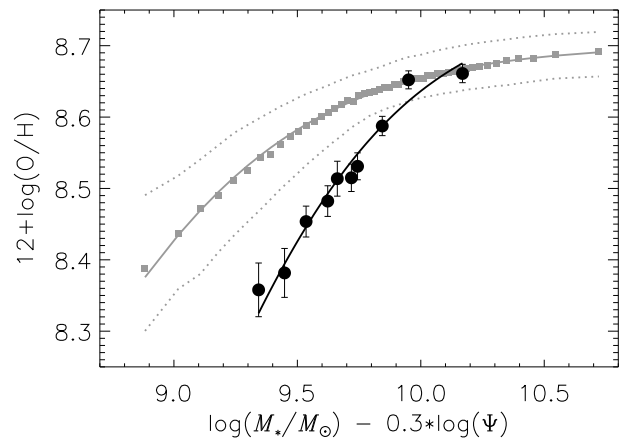}
\end{center}
\caption{The FMR for our local fiducial sample (gray curve and squares) and our $z\sim1.6$ sample (black curve and circles). The gray dotted lines denote the interval containing the central 68\% of local galaxy distribution.}
\label{fig:fmr2}
\end{figure}

\begin{figure*}[ht!]
\begin{center}
\includegraphics[width=2\columnwidth]{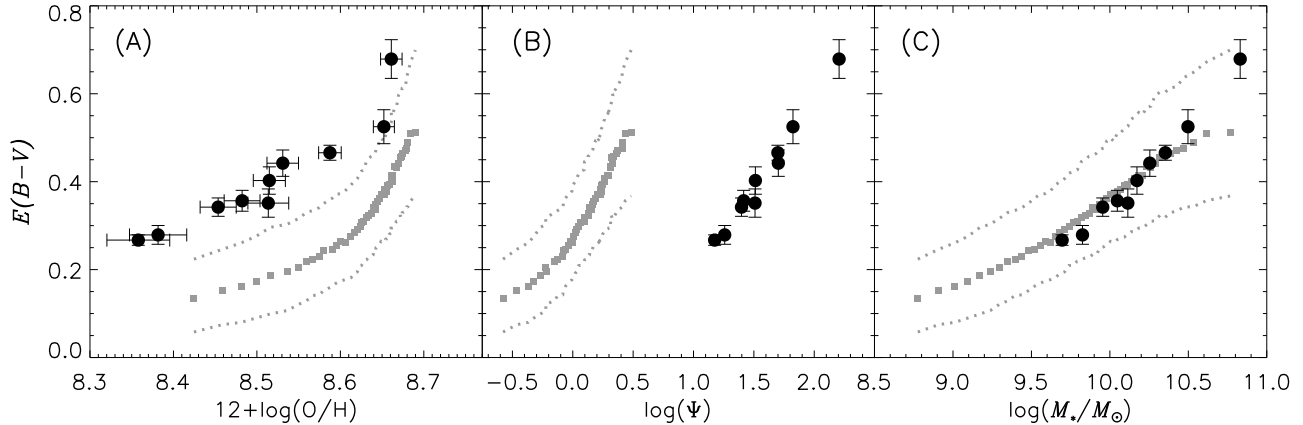}
\end{center}
\caption{The relation between $E(B-V)$ and (A) metallicity, (B) SFR and (C) stellar mass. The black points are galaxies at $z\sim1.6$ and the gray squares are the median values of $E(B-V)$ in 50 bins of stellar mass. The dotted line denotes the interval containing the central 68\% of the distribution of local galaxies.}
\label{fig:ebv_z}
\end{figure*}

The analysis of the FMR we present differs from \citet{Mannucci2010} in one significant manner. Unlike \citet{Mannucci2010} who determine SFRs from dust corrected H$\alpha$ luminosity in the fiber, we instead adopt the aperture and dust corrected SFRs provided by the MPA/JHU group in the SDSS DR7. This is very important as aperture corrections for SFRs are significant for local galaxies \citep[see][]{Zahid2013a}. For the SDSS sample used in this study the aperture corrected SFRs are on average 0.5 dex larger than SFRs determined from H$\alpha$ flux measured in the aperture. This is because the typical fiber covering fraction for galaxies in our local sample is $\sim30\%$. For the \citet{Mannucci2010} sample, the average difference is 0.7 dex. The larger difference is because they do not apply a minimum covering fraction in selecting their sample. This difference is significant and substantially impacts the FMR. The FMR with aperture corrected SFRs is offset by 0.22 dex from the FMR determined using SFRs calculated from the H$\alpha$ flux in the fiber.

If we calculate metallicities for our sample using the \citet{Maiolino2008} $N2$ calibration, which is the one used in \citet{Mannucci2010}, and apply the exact same selection and methodology in analyzing the local SDSS sample as \citet{Mannucci2010}, we find significantly better agreement between galaxies in the local Universe and at $z\sim1.6$. However, we still find that the our sample at $z\sim1.6$ is a steeper relation that is not fully consistent with the local FMR within the errors. The standard interpretation of the FMR is that it reflects short timescale responses to gas flows \citep{Mannucci2010}. Inflows of pristine gas decrease the gas-phase oxygen abundance but also lead to an increase in SFR. In this sense, it is a relation between global, integrated properties of galaxies. However, the SDSS SFRs measured without aperture corrections are not reflective of the global SFR in SDSS galaxies and therefore not the appropriate measurement for deriving relations based on global properties. It is beyond the scope of this paper to examine the FMR in detail. We simply note that if we use SDSS SFRs without aperture corrections, we find better agreement between the local FMR and the relation at $z\sim1.6$. This agreement is likely to be misleading. For the following analysis however, we use the aperture corrected SDSS SFRs.

In Figure \ref{fig:mzsfr} we examine the relation between stellar mass, metallicity and SFR. Similar to trends reported by \citet[and others]{Mannucci2010} for local SDSS galaxies, we find that at a fixed stellar mass, metallicity is \emph{anti-}correlated to the SFR. Following the approach of \citet{Mannucci2010}, we determine the value $\alpha$ that minimizes the scatter in metallicity for our $z\sim1.6$ sample. In Figure \ref{fig:fmr1} we plot the metallicity as a function $\mu_\alpha$ for galaxies shown in Figure \ref{fig:mzsfr}. Because of the small sample size, we find that the derived value of $\alpha$ is dependent on the number of bins. We conclude that for our sample of galaxies at $z\sim1.6$ the scatter is minimized for $\alpha \sim 0.4-0.5$. 

We find that for local galaxies $\alpha = 0.30$ minimizes the scatter in metallicities when they are measured using the $N2$ ratio. This same value is independently derived by \citet{Andrews2013}. In Figure \ref{fig:fmr2} we plot the metallicities of local (gray curve) and $z\sim1.6$ (black curve) galaxies as a function of the $\mu_\alpha$ that minimizes the scatter in the local relation.


Our data do not support a relation between stellar mass, metallicity and SFR that is independent of redshift, i.e. the FMR of \citet{Mannucci2010}. When the metallicities of galaxies at $z\sim1.6$ are plotted against the $\mu_\alpha$ that minimizes the scatter in the local relation, a single relation is not observed (Figure \ref{fig:fmr2}). \emph{The data support significant evolution in the relation between stellar mass, metallicity and SFR}. We emphasize that this is largely due to the use of aperture corrected SFRs for local galaxies.


\subsection{The Stellar Mass, Metallicity, SFR and Dust}

Understanding the distribution of dust as a function of cosmic time and galaxy properties is critical. Several recent studies have focused on the dust properties of local galaxies \citep[e.g.,][]{Garn2010b, Xiao2012, Zahid2013a}. \citet{Garn2010b} derive a relation between dust extinction and stellar mass. On average the magnitude of extinction, A$_{\mathrm{H}\alpha}$, varies between 0 and 2 for galaxies in the SDSS. \citet{Garn2010b} perform a principal component analysis of dust extinction, stellar mass, metallicity and SFR. From PCA, they conclude that the dominant physical property related to dust extinction in galaxies is stellar mass. The secondary correlations between dust extinction, metallicity and SFR are primarily due to the correlation of all three of these properties to stellar mass. At a fixed stellar mass dust extinction and metallicity are correlated in our sample of galaxies at $z\sim1.6$ (see Figure \ref{fig:ebv}). However, we note that this relation is significantly weaker than the relation between stellar mass and dust extinction. The straightforward interpretation is that galaxies increase their dust content as they build their stellar mass.


We examine the relation between dust extinction, stellar mass, metallicity and SFR for our sample at $z\sim1.6$ with the local relation. In Figure \ref{fig:ebv_z} we plot the dust extinction as a function of (A) metallicity, (B) SFR and (C) stellar mass. The black points are the $z\sim1.6$ sample sorted into 10 bins of stellar mass and the gray squares are the local sample sorted into 50 bins of stellar mass. Figure \ref{fig:ebv_z}A and B clearly demonstrate that dust extinction as a function of metallicity and SFR, respectively, are significantly offset from the local relation. In contrast, the relation between dust extinction and stellar mass is similar for galaxies local galaxies and $z\sim1.6$ galaxies (see also Paper I). 


The primary difference in the relation between stellar mass and dust extinction at $z\sim1.6$ as compared to the local Universe is at the high mass end. At $z\sim1.6$, massive galaxies exhibit larger extinction as compared to local galaxies. This may be due to the distribution of dust within galaxies or to a greater dust content in galaxies at $z\sim1.6$.  \citet{Wild2011} show that the line-to-continuum extinction is greater for galaxies with higher stellar mass surface densities. They interpret this as an effect related to the dust distribution around young stars. In contrast, if the dust content is greater, this may be related to the higher SFRs of galaxies at $z\sim1.6$. \citet{Zahid2013c} suggest that dust efflux by outflows may explain the distribution of dust in local star-forming galaxies. In this scenario, dust is slowly effluxed from  galaxies by continuous interaction of dust with the radiation field generated by ongoing star-formation. The timescale of dust efflux is significantly longer than the timescale of dust production and therefore galaxies accumulate dust. The dust content of galaxies in this scenario is a balance between dust production and dust efflux. A generic prediction of this model is that massive galaxies which form stars more rapidly also have greater dust content. This is because, on average, they form their stars over a shorter period of time and do not have as much time to efflux dust from their ISM.


\section{Summary and Conclusions}

We derive the mass-metallicity relation using spectroscopic observations of $\sim160$ galaxies at $z\sim1.6$. These galaxies are observed as part of our ongoing survey of star-forming galaxies in the redshift desert. These data constitute the largest high-resolution spectroscopic sample of star-forming galaxies at $z>1.4$ for which we can measure star formation rates and metallicities from optical emission lines. The main results and conclusions of our analysis are:
\begin{itemize}

\item There is a strong relation observed between stellar mass and metallicity for star-forming galaxies at $z\sim1.6$. The shape of the mass-metallicity relation evolves with redshift and is steeper at early times. The most massive galaxies ($M_\ast \sim 10^{11}M_\odot$) in our sample at $z\sim1.6$ have enriched to the level observed in the local Universe. Less massive galaxies ($M_\ast \sim 10^{9.5}M_\odot$) have metallicities that are $>0.25$ dex lower at $z\sim1.6$ as compared to the local Universe.

\item The data support our previous results showing that the evolution of the shape in the mass-metallicity relation is a consequence of galaxies enriching to an empirical upper metallicity limit. The stellar mass where galaxies enrich to this upper metallicity limit is $\sim1.2$ dex larger at $z\sim1.6$ than in the local Universe. Our analysis suggests that the upper metallicity limit does not evolve significantly since $z\sim1.6$.

\item At a fixed stellar mass, metallicity is \emph{anti}-correlated to the star formation rate such that, on average, galaxies with higher star formation rates tend to have lower metallicities. Similar trends are observed in the local Universe.

\item Our data do not support a relation between stellar mass, metallicity and SFR that is independent of redshift \citep[i.e.][]{Mannucci2010}. We observe significant evolution in the relation between stellar mass, metallicity and star formation rate when comparing the local data with our $z\sim1.6$ sample.

\item We find that at a fixed stellar mass, dustier galaxies tend to have higher metallicities. We examine the relation between dust extinction and stellar mass, metallicity and SFR for galaxies at $z\sim1.6$. By comparing these relations with the same relations for local galaxies, we conclude that stellar mass is closely related to the dust content of galaxies.

\end{itemize}

A consistent picture for the chemical evolution of star-forming galaxies since $z\sim2$ is emerging. Our analysis and conclusions are based on averaging spectra from many galaxies. Measurements of metallicities in mass-selected individual galaxy samples using multiple diagnostics will be useful for assessing systematic issues. Deeper near-infrared spectroscopic surveys with greater wavelength coverage should allow us to do this soon.

\acknowledgements We thank the anonymous referee for their careful reading and many useful suggestions that greatly improved this paper. This work is possible through the important contribution of Kentaro Aoki and the Subaru Telescope staff who assisted in acquiring much of the data presented in this work. This work was supported by World Premier International Research Center Initiative (WPI Initiative), MEXT, Japan. This work has been partially supported by the Grant-in-Aid for the Scientific Research Fund under Grant Nos. 22340056: N.S., 23224005: N.A., 25707010 and Program for Leading Graduate Schools {\it PhD Professional: Gateway to Success in Frontier Asia} commissioned by the Ministry of Education, Culture, Sports, Science and Technology (MEXT) of Japan.  We are also grateful to INAF for regular support through a grant ``PRIN-2010''. We acknowledge the importance of Mauna Kea within the indigenous Hawaiian community and with all respect say mahalo for the use of this sacred site.

\bibliographystyle{apj}
\bibliography{/Users/jabran/Documents/latex/metallicity}

 \end{document}